\documentclass[lettersize,journal]{IEEEtran}
\usepackage{array}
\usepackage[caption=false,font=normalsize,labelfont=sf,textfont=sf]{subfig}
\usepackage{stfloats}
\usepackage{url}
\usepackage{verbatim}
\def\BibTeX{{\rm B\kern-.05em{\sc i\kern-.025em b}\kern-.08em
    T\kern-.1667em\lower.7ex\hbox{E}\kern-.125emX}}
\usepackage{balance}
\usepackage{cite}
\usepackage{amsmath,amssymb,amsfonts}
\usepackage{algorithmic}
\usepackage{algorithm}
\usepackage{graphicx}
\usepackage{textcomp}
\usepackage{romannum}
\usepackage{cuted}
\usepackage{multirow}
\usepackage{diagbox}
\usepackage{array}
\usepackage{adjustbox}
\usepackage{siunitx}
\usepackage[table]{xcolor}
\usepackage{xcolor}
\usepackage{tikz}
\definecolor{CustomRed}{rgb}{1.000, 0.8906, 0.8789}
\definecolor{CustomBlue}{rgb}{0.0, 0.0, 0.5000}

\usepackage{fancyhdr}
\fancypagestyle{empty}{fancy}

\begin{document}
\title{Interference Analysis and Successive Interference Cancellation \\ for Multistatic OFDM-based ISAC Systems}
\pagenumbering{arabic}

\author{Taewon Jeong,~\IEEEmembership{Graduate Student Member,~IEEE,} Lucas Giroto,~\IEEEmembership{Member,~IEEE,} Umut Utku Erdem,~\IEEEmembership{Graduate Student Member,~IEEE,} Christian Karle,~\IEEEmembership{Graduate Student Member,~IEEE,} Jiyeon Choi,~\IEEEmembership{Graduate Student Member,~IEEE,} Thomas Zwick,~\IEEEmembership{Fellow,~IEEE,} and Benjamin Nuss,~\IEEEmembership{Senior Member,~IEEE,}

\thanks{The authors acknowledge the financial support by the Federal Ministry of Research, Technology and Space of Germany in the projects ``KOMSENS-6G'' (grant number: 16KISK123) and ``Open6GHub'' (grant number: 16KISK010). \textit{(Corresponding author: Taewon Jeong.)}}
\thanks{
Taewon Jeong, Umut Utku Erdem, Jiyeon Choi, Thomas Zwick, and Benjamin Nuss are with the Institute of Radio Frequency Engineering and Electronics (IHE), Karlsruhe Institute of Technology (KIT), 76131 Karlsruhe, Germany (e-mail: taewon.jeong@kit.edu; umut.erdem@kit.edu; jiyeon.choi@kit.edu; thomas.zwick@kit.edu; benjamin.nuss@kit.edu).
}
\thanks{Lucas Giroto was with the Institute of Radio Frequency Engineering and Electronics (IHE), Karlsruhe Institute of Technology (KIT), 76131 Karlsruhe, Germany. He is now with Nokia Bell Labs, 70469 Stuttgart, Germany (e-mail: lucas.giroto@nokia-bell-labs.com).}
\thanks{
Christian Karle is with the Institute for Information Processing Technology (ITIV),
Karlsruhe Institute of Technology (KIT), 76131 Karlsruhe, Germany (e-mail: christian.karle@kit.edu).
}}

\newcommand{\salmonline}{\raisebox{2pt}{\tikz{\draw[dashed, line width=1.5pt, color={rgb:red,0.888; green,0.154; blue,0.154}] (0,0) -- (6mm,0);}}}
\newcommand{\greenline}{\raisebox{2pt}{\tikz{\draw[dashed, line width=1.5pt, color={rgb:red,0.154; green,0.888; blue,0.154}] (0,0) -- (6mm,0);}}}
\newcommand{\blueline}{\raisebox{2pt}{\tikz{\draw[dashed, line width=1.5pt, color={rgb:red,0.154; green,0.154; blue,0.888}] (0,0) -- (6mm,0);}}}

\maketitle

\begin{abstract}
Multistatic integrated sensing and communications (ISAC) systems, which use distributed transmitters and receivers, offer enhanced spatial coverage and sensing accuracy compared to stand-alone ISAC configurations. However, these systems face challenges due to interference between co-existing ISAC nodes, especially during simultaneous operation. In this paper, we analyze the impact of this mutual interference arising from the co-existence in a multistatic ISAC scenario, where a mono- and a bistatic ISAC system share the same spectral resources. We first classify differenct types of interference in the power domain. Then, we discuss how the interference can affect both sensing and communications in terms of bit error rate (BER), error vector magnitude (EVM), and radar image under varied transmit power and RCS configurations through simulations. Along with interfernce analysis, we propose a low-complexity successive interference cancellation method that adaptively cancels either the monostatic reflection or the bistatic line-of-sight signal based on a monostatic radar image signal-to-interference-plus-noise ratio (SINR). The proposed framework is evaluated with both simulations and proof-of-concept measurements using an ISAC testbed with a radar echo generator for object emulation. The results have shown that the proposed method reduces BER and improves EVM as well as radar image SINR across a wide range of SINR conditions. These results demonstrate that accurate component-wise cancellation can be achieved with low computational overhead, making the method suitable for practical applications.
\end{abstract}


\section{Introduction}
\IEEEPARstart{W}{ith} the rapid advancement of wireless communication technologies, spectrum scarcity has become a major challenge, particularly for fifth-generation (5G) and upcoming sixth-generation (6G) cellular systems. To address this problem, integrated sensing and communications (ISAC) has been proposed as a solution that enables the use of a single spectrum resource for both communications and sensing. Recent researches have investigated the technological development of ISAC and its potential applications in future wireless networks \cite{Survey1, Survey2, Survey3, Survey4}. In response to this growing interest, the 3rd Generation Partnership Project (3GPP) has initiated standardization activities for ISAC in 6G communications. In addition, the European Telecommunications Standards Institute has specified a wide range of use cases for ISAC systems \cite{ETSI_usecase}. Typical examples of ISAC applications include drone detection \cite{Usecase_drone1, Usecase_drone2, Usecase_drone3, Usecase_drone4}, traffic and road monitoring \cite{Usecase_road1, Usecase_road2, Usecase_road3}, and target localization \cite{Usecase_loc1, Usecase_loc2, Usecase_loc3}.  

While recent studies have primarily focused on the development of ISAC technologies and their potential applications, most of the existing researches have focused on waveform design, beamforming optimization, resource management, and MIMO applicability for stand-alone ISAC systems, where a single transceiver performs both transmission and reception for communication and sensing tasks \cite{Waveform_design1, Waveform_design2, Beamforming_optimize1, Beamforming_optimize2, Resource_management, MIMO_ISAC}. Although stand-alone ISAC systems suffer from self-interference, which is difficult to suppress and can limit sensing accuracy, this configuration simplifies system design and synchronization requirements, making it a practical option for the first phase of ISAC implementations. However, in real-world scenarios, extending ISAC systems to bi- and multistatic configurations is essential to achieve enhanced sensing accuracy, wider coverage, and improved robustness against blockage effects. In particular, multistatic ISAC systems, where multiple transmitters (Tx) and receivers (Rx) cooperate in a distributed manner, offer the potential for high-resolution sensing and reliable communications in complex environments. While these configurations are important, research on bi- and multistatic ISAC systems have recently gained attention. For example, the authors in \cite{Bistatic_1} proposed a system architecture and over-the-air synchronization method for bistatic ISAC, whereas \cite{Bistatic_2} presented a beamforming framework tailored for bistatic MIMO ISAC systems. Additionally, an artificial intelligence-based approach for angle-of-arrival and angle-of-departure estimation is introduced in \cite{Bistatic_DOA}. Research efforts targeting multistatic ISAC systems and cooperative ISAC networks have also emerged, as can be seen in \cite{Multistatic_1, Multistatic_2, CoMP}. Despite these initial developments, the exploration of practical bi- and multistatic ISAC systems remains in its early stages, with many challenges to address for their widespread adoption.

\begin{figure*}[!t]
	\begin{center}
		\includegraphics[width = 1\linewidth]{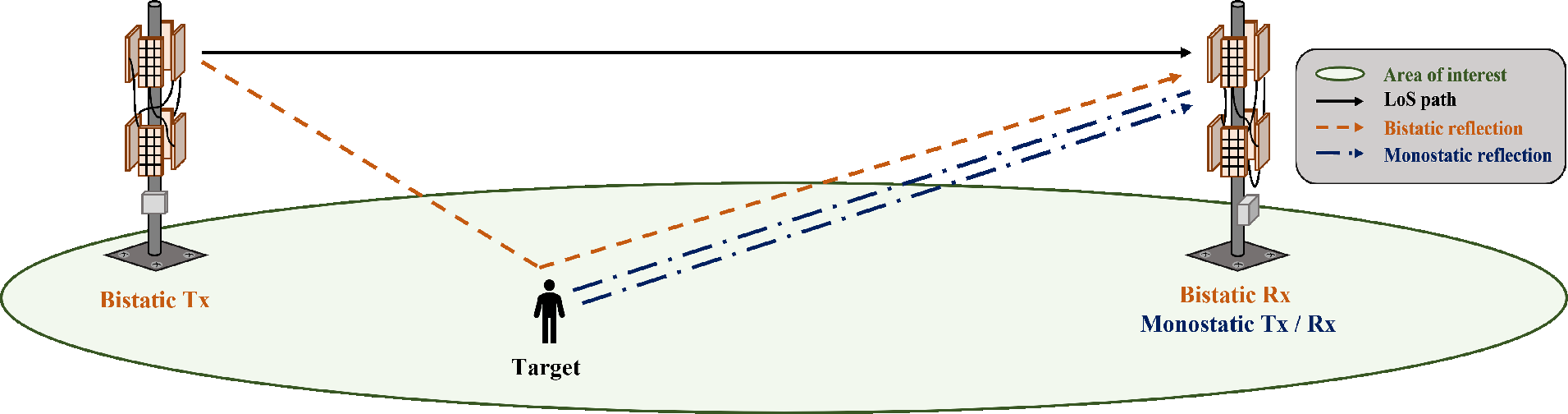}
		\caption{Representation of the multistatic ISAC system where one node serves as bistatic transmitter and the other node serves as monostatic transceiver and bistatic receiver.}\label{Fig1}
	\end{center}
\end{figure*}

In multistatic ISAC networks, one of the most critical challenges is the management and cancellation of interference between distributed ISAC nodes. Since multiple transmitters and receivers are spatially deployed across the coverage area, interference between ISAC nodes is inevitable. Several studies have investigated the types of interference that arise in multistatic ISAC systems and proposed various interference avoidance techniques \cite{Interference_survey1, Interference_survey2, Interference_survey3,Interference_survey4}. In addition, various methods w.r.t. interference management have been proposed which are based on precoding, antenna design, beamforming, and resource management strategies, such as time-division duplexing \cite{Interference_Precoding1, Interference_beamforming, Interference_TDD}. Along with interference management, interference cancellation techniques have also been proposed in \cite{Interference_cancellation1, Interference_cancellation2, Interference_cancellation3, Interference_cancellation4}. While the aforementioned methods showed promising performance regarding interference management and cancellation, it is true that the validation in real scenarios are lacking. Moreover, problems regarding computational burden may arise in real-time processing since cancellation is done with optimization and iterative process. Together with those limitations, to the best of our knowledge, there is yet no study which analyzes the effect of interference in terms of both sensing and communications in ISAC systems. Furthermore, most of the prior works focus on mitigating a specific type of interference (e.g., self-interference) while leaving other types of interference unsolved.

In this paper, we analyze the impact of interference between ISAC nodes in multistatic ISAC network. Specifically, we consider a scenario where one monostatic ISAC system and one bistatic ISAC system co-exist within the region of interest, as illustrated in Fig. \ref{Fig1}. The interference is analyzed for all possible combinations of communication and sensing operations, including monostatic sensing, bistatic communication, and bistatic sensing. For the interference analysis, we define the signal-to-interference-plus-noise ratio (SINR), which is calculated as the ratio of the received power of the overall bistatic signal — consisting of both bistatic communication and bistatic sensing signals — to the power of the monostatic reflection and thermal noise at the receiver. The detailed SINR formulation is provided in Sec. \ref{Sec2}. Together with the interference analysis, we propose a successive interference cancellation (SIC) method that utilizes the estimated radar and communication channel information. The proposed SIC method adaptively cancels either the monostatic reflection or the bistatic line-of-sight (LoS) component, depending on the operating scenario, to improve both communication and sensing performances. Moreover, the proposed cancellation method requires low computational complexity, making it suitable for practical implementation. Finally, the effectiveness of the proposed interference analysis and cancellation method is validated through numerous simulations, and also through the proof-of-concept measurements.

In summary, major contributions of this paper can be summarized as follows:
\begin{itemize}
	\item We analyze various types of interference in multistatic ISAC systems, including self-interference and mutual interference between ISAC nodes. Also, their impact on both sensing and communication performance is quantitatively evaluated through simulations.

	\item We propose a low-complexity SIC method that mitigates mono- and bistatic interference. The proposed method demonstrates comparable cancellation performance to existing approaches while maintaining low computational overhead. Furthermore, its effectiveness is validated through proof-of-concept measurements with ISAC testbed at \qty{3.68}{GHz} carrier frequency.
\end{itemize}

The remainder of this paper is organized as follows. Sec. \ref{Sec2} presents the system model, detailed mathematical formulation of monostatic and bistatic ISAC signals, and analysis of  inteferences. Sec. \ref{Sec3} introduces the proposed successive interference cancellation method, it's result with numerous simulations, and derivation of computational complexity. Then, Sec. \ref{Sec4} shows the proof-of-concept measurement setup and its results. Finally, we conclude the paper and outline potential future research in Sec. \ref{Sec5}.

\section{Signal Model at Multistatic ISAC Receiver}\label{Sec2}
In this section, we begin by presenting the different types of ISAC-related signal components observed at the receiver, followed by the mathematical model of the received signal. Based on this model, each individual signal component is classified in the power domain, listed from the dominant to the least dominant. This classification is crucial for the proposed SIC method that will be discussed in Sec. \ref{Sec3}. Finally, we then analyze the monostatic to bistatic interference and bistatic to monostatic interference w.r.t. both sensing and communications. For simplicity of the analysis, we consider a scenario with a single point target and two ISAC nodes within the region of interest: one acting as a bistatic transmitter, and the other serving as both a bistatic receiver and a monostatic transceiver, as illustrated in Fig.~\ref{Fig1}.

\subsection{System Model}\label{Sec2_1}

Let $\mathbf{T}_{\text{mono}} \in \mathbb{C}^{N \times M}$ and $\mathbf{T}_{\text{bi}} \in \mathbb{C}^{N \times M}$ be the transmitted frame from each monostatic and bistatic ISAC transmitter. 
Each $N$ and $M$ denotes the total number of subcarriers and symbols that make up the single orthogonal frequency-division multiplexing (OFDM) frame, respectively. 
Data modulation in OFDM-based communication systems takes place in the frequency domain, and an inverse discrete Fourier transform (IDFT) is applied to obtain the discrete time-domain signal, which can be expressed as
\begin{align}\label{Eq1}
	x_m[k] = \frac{1}{N} \sum_{n=0}^{N-1} X_m[n] \, \mathrm{e}^{j 2 \pi \frac{n k}{N}}.
\end{align}
\begin{figure*}[!h]
	\begin{equation}\label{Eq2}
		y(t) = \left(s_{\mathrm{SI}}(t) + \sum_{l=0}^{L-1}\alpha_l s_{\mathrm{mono}}(t-\tau_l)e^{j2\pi f_{D_{\mathrm{mono}},l}t}\right) + \left(\alpha_{\mathrm{LoS}}s_{\mathrm{bi}}(t-\tau_{\mathrm{LoS}}) + \sum_{p=1}^{P-1}\alpha_p s_{\mathrm{bi}}(t-\tau_p)e^{j2\pi f_{\mathrm{D_{bi}},p}t}\right) + w(t)
	\end{equation}
	\hrulefill
	\vspace*{6pt}
\end{figure*}
In \eqref{Eq1}, $n \in \{0, 1, ..., N-1\}$ denotes the subcarrier index, and $k \in \{0, 1, ..., N-1\}$ denotes the index of discrete time-domain samples. After converting the signal to time-domain, $N_{\text{cp}}$-length of cyclic prefix (CP) is prepended in the front of the signal to avoid inter-symbol interference (ISI). Converting to the time-domain and prepending the CP is done throughout all symbols in one OFDM frame, resulting in the matrix $\mathbf{X}_{\text{mono}} \in \mathbb{C}^{(N_{\text{cp}} + N) \times M}$. Next, $\mathbf{X}_{\text{mono}}$ undergoes parallel-to-serial (P/S) conversion. Then the in-phase ($\mathrm{I}$) and quadrature ($\mathrm{Q}$) component of the output signal is fed into the digital-to-analog converter, each modulated by $\text{cos}(2\pi f_{\text{c}} t)$ and $\text{sin}(2\pi f_{\text{c}} t)$, where $f_{\text{c}}$ is the carrier frequency. The two resulting signals are summed and transmitted with the power of $P_{\text{Tx}}$ and the antenna gain of $G_{\text{Tx}}$. 

\begin{figure}[!b]
	\begin{center}
		\includegraphics[width = 1\linewidth]{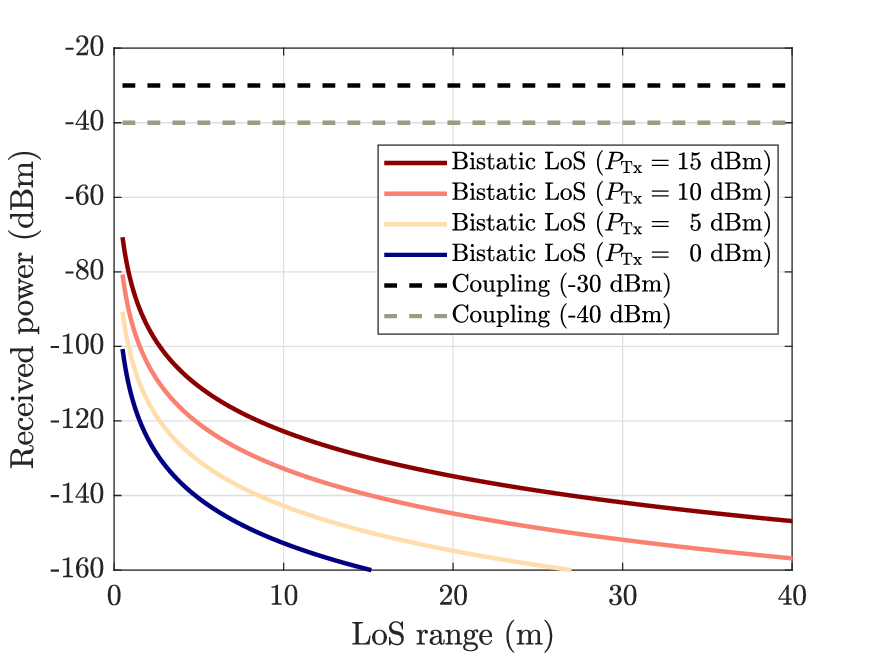}
		\caption{Comparison of received power between direct coupling and bistatic LoS path.}\label{Fig2}
	\end{center}
\end{figure}

Assuming the channel-induced received signal as $r(t)$, the first step at the receiver is to undergo low-noise amplifier (LNA), band-pass filter (BPF), and the down-conversion to the baseband (BB). 
The down-conversion is performed by mixing $r(t)$ with $\text{e}^{-\text{j} 2 \pi f_{\text{c}} t}$ and applying a low-pass filter to eliminate the high-frequency components generated during mixing, thereby yielding the complex baseband signal. Once the down-conversion is completed, the down-converted signal undergoes analog-to-digital conversion (ADC), where the continuous-time baseband signal is sampled and quantized. This results in discrete-time in-phase ($\mathrm{I}$) and quadrature ($\mathrm{Q}$) components. To construct the complex digital baseband signal, the $\mathrm{Q}$ component is multiplied by the imaginary unit $\text{j}$ and added to the $\mathrm{I}$ component. The resulting complex signal is then passed through a serial-to-parallel (S/P) converter. Following the S/P conversion, the CP, which was inserted at the transmitter side is removed from the received signal, resulting in the matrix $\mathbf{Z}_{\text{mono}} \in \mathbb{C}^{N \times M}$. This CP-free matrix is then used as the input for radar signal processing or communication processing steps. Although the described signal processing steps are explained in the context of monostatic transmission and reception, the same processing chain is applied to the bistatic ISAC system as well. The key difference, however, lies in the synchronization requirements, as the transmitter and receiver operate with separate hardware and the receiver lacks knowledge of the transmitted symbols. In particular, the unknown arrival time of the received signal introduces a symbol timing offset (STO) at the receiver. In addition, mismatches in the carrier frequency and sampling clocks between the transmitter and receiver, caused by the use of independent local oscillators and sampling clocks, introduce additional impairments such as carrier frequency offset (CFO) and sampling frequency offset (SFO). Among various synchronization offsets, the estimation of STO and CFO is performed using the Schmidl \& Cox (S\&C) \cite{S&C}, while the estimation of SFO is carried out using the TITO algorithm proposed in \cite{TITO}.

\subsection{Power Domain Classification}\label{Sec2_2}

As shown in Fig.~\ref{Fig1}, various signals are combined at the receiver, which are LoS communication signal from bistatic transmitter, bistatic sensing reflection, and monostatic sensing reflection. Furthermore, while it is not depicted in Fig.~\ref{Fig1}, there is also a Tx-Rx spillover, so called direct coupling or self-interference due to the co-location of Tx and Rx in the monostatic system denoted as $s_{\text{SI}}(t)$. Therefore, the time-domain signal at the receiver can be expressed as $y(t)$ in \eqref{Eq2}. In \eqref{Eq2}, the two components in the first brackets on the right-hand side correspond to the signal originating from monostatic Tx. Each term $L$, $\alpha_l$, $s_{\mathrm{mono}}(t)$, $\tau_l$, $f_{\mathrm{D_{mono}},\,l}$ denotes the number of targets within the coverage, the signal attenuation factor, the transmitted waveform, the round-trip propagation delay, and the relative Doppler shift from the $l$th target, all w.r.t. the monostatic system. Next, the third and fourth terms correspond to the signals originating from the bistatic transmitter. In this expression, $P$, $\alpha_{\mathrm{LoS}}$, and $\alpha_p$ represent the total number of targets in the bistatic coverage area, the attenuation factors of the bistatic LoS path and the $p$th target reflection, respectively. The transmitted signal is denoted by $s_{\mathrm{bi}}(t)$. The delays $\tau_{\mathrm{LoS}}$ and $\tau_{p}$ represent the one-way and two-way propagation delays associated with the LoS path and target reflection, respectively. Similarly, $f_{\mathrm{D_{bi}},p}$ denote the relative Doppler shifts from the $p$th target path. Lastly, the term $w(t)$ denotes the additive white Gaussian noise.

When excluding the noise term on the right-hand side of \eqref{Eq2}, the signal component with the highest power is typically the $s_{\text{SI}}(t)$ that occurs at the monostatic radio frequency (RF) frontend. Fig.~\ref{Fig2} illustrates the received power of the bistatic LoS path as a function of range, assuming a constant transmit power at the monostatic transmitter. For reference, two typical coupling levels at \qty{-30}{dBm} and \qty{-40}{dBm} are shown to indicate representative coupling powers relative to the transmit power \cite{Coupling}. Here, the bistatic transmit power $P_{\text{Tx}}$ varies from \qty{0}{dBm} to \qty{15}{dBm}, while the antenna gain is fixed at \qty{10}{dBi} for both the transmit and receive sides, and the carrier frequency is set to \qty{28}{GHz}. Whereas the assumed power levels may differ from those used in standard-compliant communication systems (e.g., \qty{46}{dBm} transmit power at 5G gNodeB Tx \cite{gNB_Ptx}), they were selected to align with the measurement conditions presented later in Sec.~\ref{Sec4}. As shown in Fig.~\ref{Fig2}, the received power of the LoS path remains consistently lower than all coupling levels across the considered distances and transmit power values. Even at short distances such as \qty{1}{m} and with the highest transmit power of \qty{15}{dBm}, the LoS signal does not exceed the coupling power. This indicates that the direct coupling remains as the dominant signal component at the receiver unless additional suppression is applied.

Following the direct coupling, the bistatic LoS signal is the next strongest component in general. This is because the LoS communication path undergoes only one-way propagation loss, unlike mono- and bistatic reflections, which involve two-way propagation. The received power of the bistatic LoS can be expressed as
\begin{align}\label{Eq3}
	P_{\text{bi,LoS}} = \frac{P_{\text{Tx}} G_{\text{Tx}} G_{\text{Rx}} \lambda_{\text{c}}^2}{(4\pi R_{\text{LoS}})^2}.
\end{align}
Also, the power of both mono- and bistatic reflection can be expressed as
\begin{align}\label{Eq4}
	P_{\text{mono,tgt}} = \frac{P_{\text{Tx}} G_{\text{Tx}} G_{\text{Rx}} \lambda_{\text{c}}^2 \sigma_{\text{mono}}}{(4\pi)^3 R^4}
\end{align}
and
\begin{align}\label{Eq5}
	P_{\text{bi,tgt}} = \frac{P_{\text{Tx}} G_{\text{Tx}} G_{\text{Rx}} \lambda_{\text{c}}^2 \sigma_{\text{bi}}}{(4\pi)^3 R_{\text{Tx}\rightarrow \text{Tgt}}^2 R_{\text{Tgt}\rightarrow \text{Rx}}^2}.
\end{align}
For notational simplicity, we assume that the transmit power and antenna gains denoted as  $P_{\text{Tx}}$, $G_{\text{Tx}}$, and $G_{\text{Rx}}$, are equal for both the mono- and bistatic ISAC nodes. The terms $\lambda_{\text{c}}$, $\sigma_{\text{mono}}$, and $\sigma_{\text{bi}}$ represent the carrier wavelength, monostatic radar cross section (RCS), and bistatic RCS, respectively. In addition, $R_{\text{LoS}}$, $R_{\text{Tx}\rightarrow \text{Tgt}}$, $R_{\text{Tgt}\rightarrow \text{Rx}}$, and $R$ denote the distance between the bistatic transmitter and receiver, the distance from the bistatic transmitter to the target, the distance from the target to the bistatic receiver, and the monostatic range, respectively. To compare the received powers of the bistatic LoS path and monostatic target reflection, the power ratio can be written as
\begin{align}\label{Eq6}
	\frac{P_{\text{bi,LoS}}}{P_{\text{mono,tgt}}} = \frac{4\pi R^4}{\sigma_{\text{mono}} R_{\text{LoS}}^2 }.
\end{align}
Under the assumption where $R^4> R_{\text{LoS}}^2 \sigma_{\text{mono}}$ is satisfied in \eqref{Eq6}, the power of the bistatic LoS path will always be at least \qty{11}{dB} higher than that of the monostatic reflection. However, in cases where the target is located near the monostatic transceiver or exhibits extremely high reflectivity, the power of monostatic reflection may become comparable to that of the bistatic LoS path. Similarly, the power of the bistatic LoS path will always be higher than that of the bistatic reflection under the assumption that
\begin{align}\label{Eq7}
	\left(4 \pi R_{\text{Tx}\rightarrow \text{Tgt}} R_{\text{Tgt}\rightarrow \text{Rx}}\right)^2> R_{\text{LoS}}^2 \sigma_{\text{bi}}.
\end{align}

\definecolor{Silver}{rgb}{0.888,0.888,0.888}
\begin{table*}[!h]
	\centering
	\renewcommand{\arraystretch}{1.4}
	\caption{Adopted system parameters and resulting performance metrics for the proposed OFDM-based ISAC framework.}
	\label{Tab1}
	\begin{tabular}{l|l||c|l|c}
		\hline \hline 
		\rowcolor{Silver}
		\multicolumn{2}{c||}{\textbf{OFDM system parameters}} & \multicolumn{3}{c}{\textbf{Performance parameters}} \\ 
		\hline \hline
		Carrier frequency & 28 GHz & \multirow{5}{*}{\textbf{Monostatic}} & Range resolution & 0.3 m \\
		Bandwidth & 500 MHz & & Maximum unambiguous range & 613.97 m \\
		Number of subcarriers & 2048 & & Doppler resolution & 381.46 Hz \\
		Number of symbols & 512 & & Maximum unambiguous Doppler & 97.65 kHz \\
		Cyclic prefix length & 512 & & Processing gain (full-frame) & 60.21 dB \\
		\cline{3-5}
		Subcarrier spacing & 244 kHz & \multirow{6}{*}{\textbf{Bistatic}} & Data rate (100\% duty cycle) & 0.6 Gbps \\
		Sampling interval & 2.0 ns & & Range resolution & 0.6 m \\
		OFDM frame duration & 2.6 ms & & Maximum unambiguous range & 1227.9 m \\
		Pilot type & Lattice & & Doppler resolution & 381.46 Hz \\
		Pilot interval (both subcarrier and symbol axes) & 2,\,2 & & Maximum unambiguous Doppler & 97.65 kHz \\
		Modulation type & QPSK & & Processing gain (full-frame) & 60.21 dB \\
		\hline
	\end{tabular}
\end{table*}

\begin{figure*}[!t]
	\begin{center}
		\includegraphics[width = 0.32\linewidth]{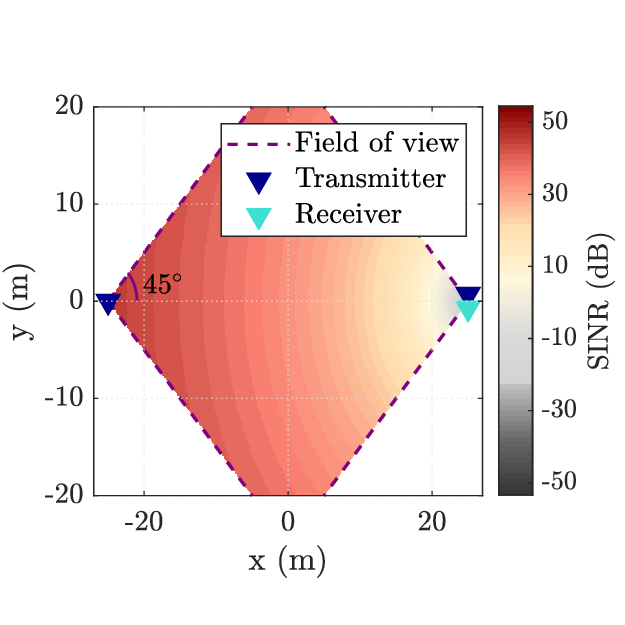}
		\includegraphics[width = 0.32\linewidth]{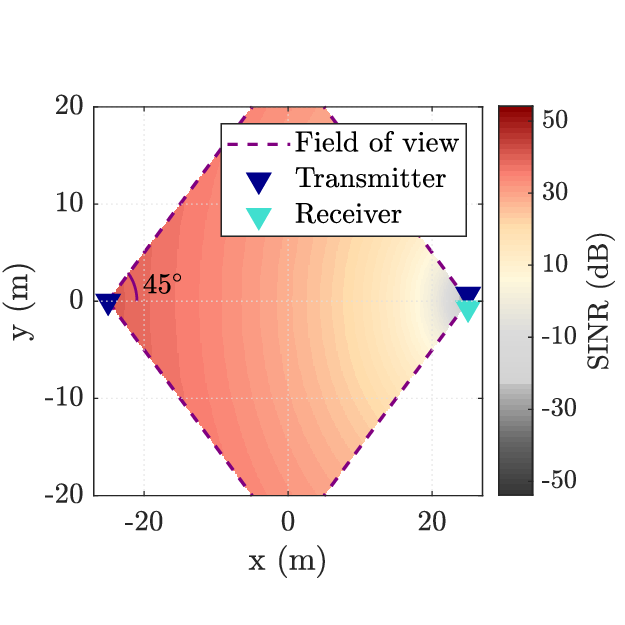} 
		\includegraphics[width = 0.32\linewidth]{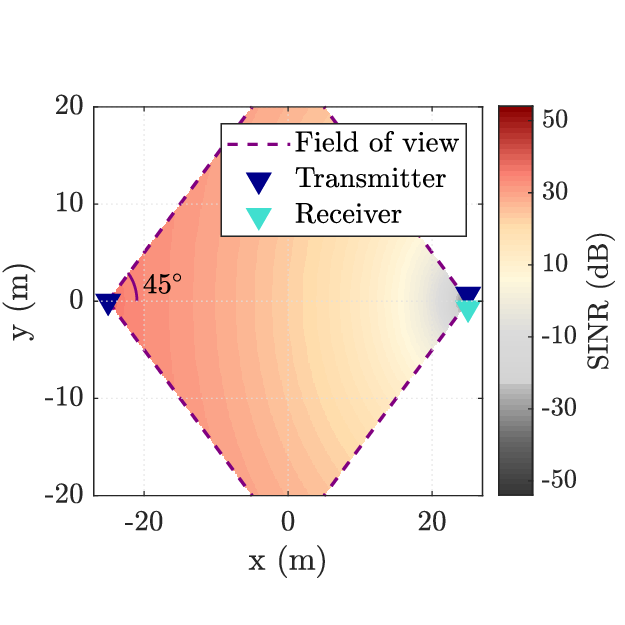}  
		\\ (a) \qquad\qquad\qquad\qquad\qquad\qquad\qquad\qquad (b) \qquad\qquad\qquad\qquad\qquad\qquad\qquad\qquad (c)
		\caption{Spatial SINR distribution across the 2D plane for different target RCS values: (a) \qty{0}{dBsm}, (b) \qty{5}{dBsm}, (c) \qty{10}{dBsm}.}\label{Fig3}
	\end{center}
\end{figure*}

Lastly, we compare the power of monostatic reflection and bistatic reflection. By taking the ratio between $P_{\text{bi,tgt}}$ and $P_{\text{mono,tgt}}$, we obtain  
\begin{align}\label{Eq8}
	\frac{P_{\text{bi,tgt}}}{P_{\text{mono,tgt}}} = \left(\frac{R^4}{R_{\text{Tx}\rightarrow \text{Tgt}}^2 R_{\text{Tgt}\rightarrow \text{Rx}}^2}\right)\left(\frac{\sigma_{\text{bi}}}{\sigma_{\text{mono}}}\right).
\end{align}
However, further analysis of this ratio is highly challenging due to its strong dependency on the specific scenario. The power comparison between mono- and bistatic reflections is influenced not only by geometry (i.e., the propagation ranges), but also by the RCS, which itself varies with several physical parameters such as material reflectivity, wavelength, incident and reflection angles, target shape, and polarization. While some studies have proposed the methods to approximate bistatic RCS from monostatic RCS, these approaches are typically suited for long-range scenarios, such as aircraft detection \cite{RCS1, RCS2}. Therefore, a generalized comparison between monostatic and bistatic reflections in the power domain is beyond the scope of this work and thus not considered in this paper.

\begin{figure*}[!t]
	\begin{center}
		\includegraphics[width = 1\linewidth]{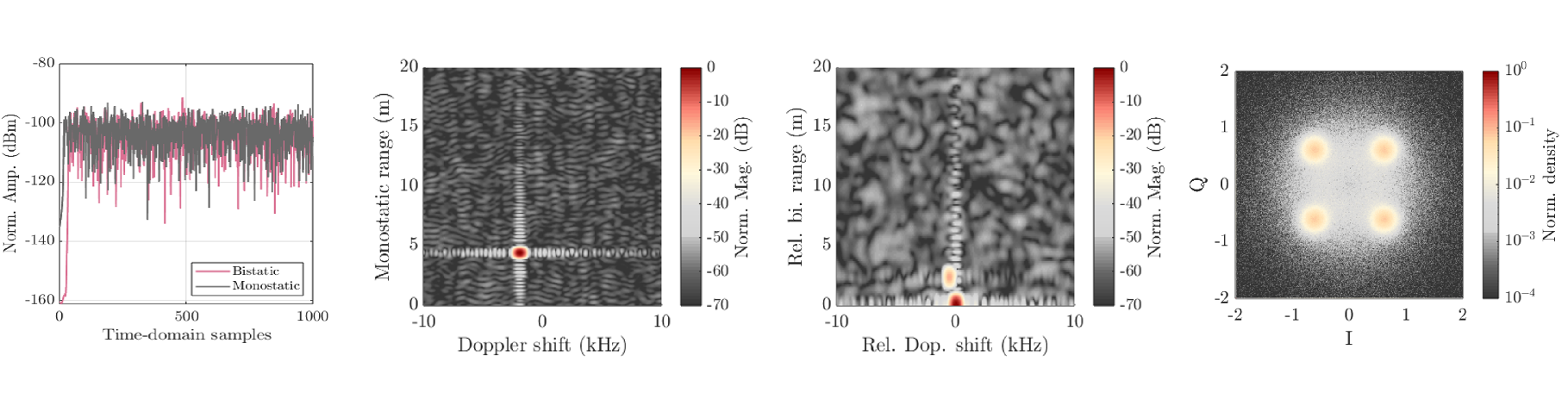}
		\\ (a) \\
		\includegraphics[width = 1\linewidth]{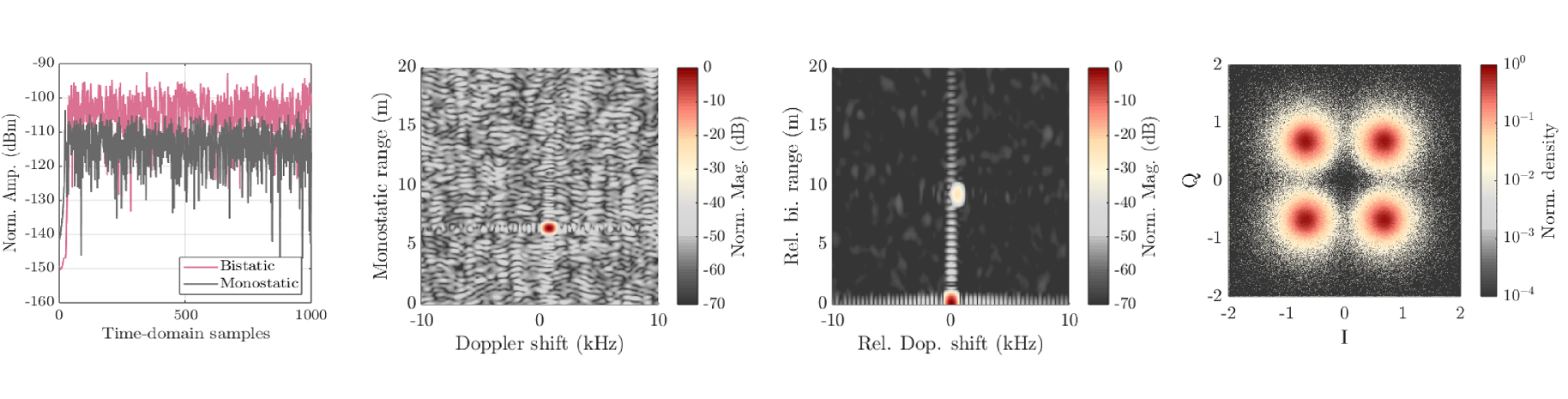}
		\\ (b) \\
		\includegraphics[width = 1\linewidth]{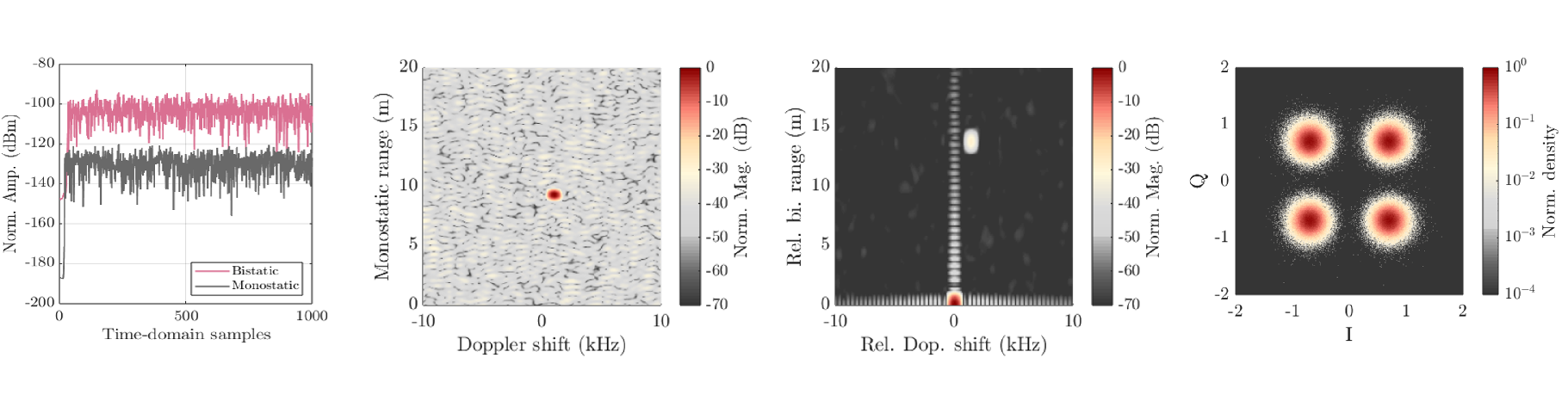} 
		\\ (c)\\
		\caption{Sensing and communication performance under interference-induced conditions for varied SINR levels. From the first row, each row corresponds to \qty{0}{dB}, \qty{10}{dB}, \qty{20}{dB} SINR, respectively. All radar images are individually normalized to their respective peak amplitudes and the constellation plot represents the normalized density in the complex plane, where denser symbol clusters indicate lower error rates.}\label{Fig4}
	\end{center}
\end{figure*}

\subsection{Interference Analysis}\label{Sec2_3}
To evaluate the impact of interference on both sensing and communication performances, we consider a simulation scenario identical to that illustrated in Fig.~\ref{Fig1}. It is important to note that the direct coupling is assumed to be canceled by the proposed cancellation technique, which will be introduced in the following section. Without this assumption, the Tx-Rx coupling would dominate the received signal unless the bistatic transmitter operates with significantly higher transmit power. Such conditions would bias a fair analysis of the interference between sensing and communication signals in a multistatic ISAC system. Therefore, the direct coupling is excluded in interference analysis. In addition, we assume the worst-case scenario, where the bistatic signal fully overlaps with the monostatic signal in the time-domain or vise versa. A single-input single-output antenna configuration is considered and detailed simulation parameters are summarized in Table~\ref{Tab1}. While these parameters may not completely comply with standard numerologies such as those defined in 5G new radio, they were intentionally chosen to align with the measurement setup described in Sec.~\ref{Sec4}.

We begin by defining the SINR that is used throughout in this work, which can be expressed as
\begin{align}\label{Eq10}
	\mathrm{SINR} = \frac{\left(P_{\text{bi,LoS}} + P_{\text{bi,tgt}}\right)}{\left(P_{\text{mono,tgt}} + P_{\text{noise}}\right)},
\end{align}
where $P_{\text{bi,LoS}}$, $P_{\text{bi,tgt}}$, $P_{\text{mono,tgt}}$, and $P_{\text{noise}}$ denote the power of LoS signal, bistatic reflection, monostatic reflection and white noise, respectively. SINR at the receiver can vary depending on several factors. Disparities in system parameters for both mono- and bistatic configuarations, such as transmit power, can affect the received power levels. Additionally, the SINR is influenced by the target’s RCS and its spatial position, which determine the strength of the reflected signal. Fig.~\ref{Fig3} presents the spatial distribution of SINR for a single target positioned at various locations on a two-dimensional (2D) plane, assuming a fixed transmit power of \qty{10}{dBm}. In each case, the SINR is calculated at the receiver based on the target’s position and RCS. For example, when the target is located at the origin in Fig.~\ref{Fig3}(c), the resulting SINR is approximately \qty{13.14}{dB}. However, the absolute values of the transmit power or RCS are not the main determinants of interference effects. Rather, it is the overall SINR at the receiver that ultimately determines the system performance, regardless of whether the interference arises from differences in system hardware configurations or from target-related characteristics. Accordingly, we adopt a simplified yet representative simulation scenario in which the monostatic and bistatic configurations share identical system parameters and RCS values. While RCS is inherently scenario-dependent, as discussed in Section~\ref{Sec2_2}, it scales the received signal power and does not affect the underlying interference dynamics. Therefore, this simplification does not compromise the generality of the results.

Fig.~\ref{Fig4} presents the effect of interference based on simulation with varied SINR values, specifically SINR equal to \qty{0}{dB}, \qty{10}{dB}, and \qty{20}{dB} respectively. For all three cases, transmit power and target RCS are fixed at \qty{10}{dBm} and \qty{5}{dBsm}. The leftmost column of Fig.~\ref{Fig4} shows the first 1,000 samples of the time-domain received signal, showing both the monostatic reflection and the overall bistatic signal. The second and third columns show the corresponding monostatic and bistatic radar images, while the rightmost column shows the demodulated constellation diagrams. As shown in the first row (i.e., \qty{0}{dB} SINR), the power of the bistatic signal is nearly comparable to that of the monostatic reflection. Under this condition, the LoS communication signal is not strong enough to support reliable communication, resulting in a severely distorted constellation and a bit error rate (BER) of approximately 0.11. It is worth emphasizing that, particularly in ISAC systems, even a small number of bit errors may degrade the bistatic radar imaging performance \cite{Bistatic_David}. While the genie-aided approach used in \cite{genie1} and \cite{genie2} can provide the full knowledge of symbols, we did not assume such idealized approach in this work. Therefore, the bistatic radar image in this case shows substantial noise artifacts due to the imperfect communication link. At \qty{10}{dB} SINR, the bistatic signal becomes stronger relative to the monostatic reflection, leading to improved but not yet error-free communication. The BER decreases to $8 \times 10^{-4}$. Compared to the \qty{0}{dB} SINR case, the monostatic radar image shows a decrease in image SINR due to reduced relative power, while the bistatic image SINR improves as interference from the monostatic component becomes less significant. It is important to note that image SINR is different from SINR defined in \eqref{Eq10}. While the latter refers to the power ratio between desired and interfering signals in the time-domain, the image SINR is defined as the ratio between the target peak and the overall noise floor in the radar image. At \qty{20}{dB} SINR, the bistatic signal dominates the monostatic signal, enabling error-free communication. However, this results in further reduction in the monostatic image SINR, as the monostatic signal becomes increasingly weak. In contrast, the bistatic radar image quality continues to improve.

\begin{figure*}[!t]
	\begin{center}
		\includegraphics[width = 0.32\linewidth]{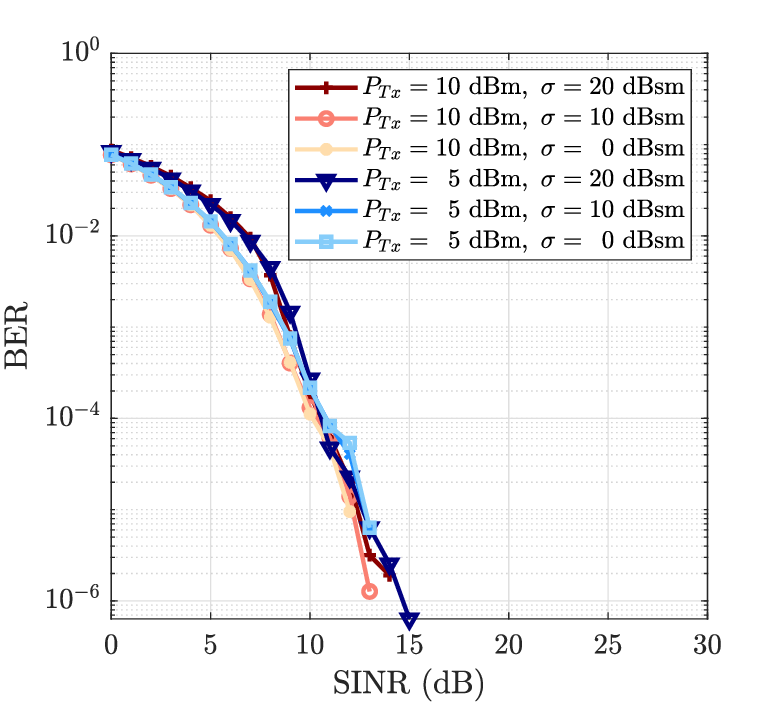}
		\includegraphics[width = 0.32\linewidth]{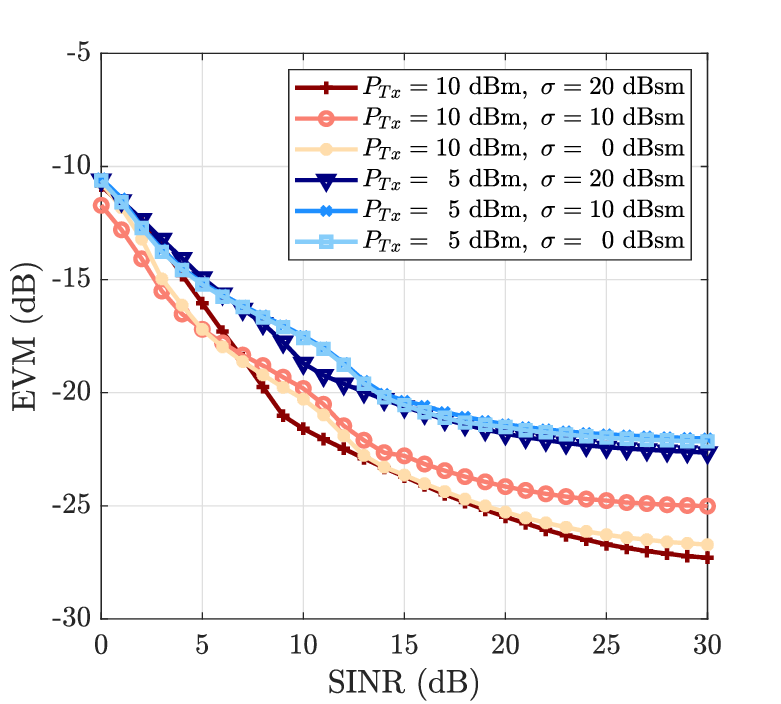} 
		\includegraphics[width = 0.32\linewidth]{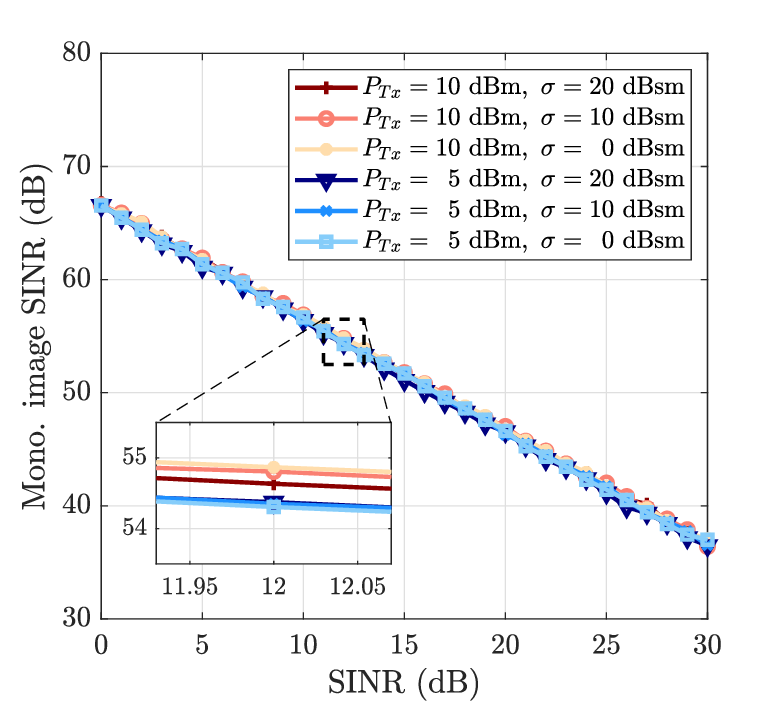}  
		\\ (a) \qquad\qquad\qquad\qquad\qquad\qquad\qquad\qquad (b) \qquad\qquad\qquad\qquad\qquad\qquad\qquad\qquad (c)
		\caption{Communication and sensing performance comparison with different transmit power and RCS at different SINR values: (a) BER, (b) EVM, (c) monostatic radar image SINR.}\label{Fig5}
	\end{center}
\end{figure*}

Moreover, simulations were conducted under a range of SINR conditions to analyze the effect of interference on both sensing and communication performances. By amplifing or attenuating the monostatic-oriented signal, the SINR was swept from 0 dB to 30 dB. In addition, a total of six representative cases were considered by varying the transmit power and RCS values, specifically $(P_{\text{Tx}}, \sigma) =$ (10, 20), (10, 10), (10, 0), (5, 20), (5, 10), and (5, 0), expressed in the units of dBm and dBsm, respectively. Fig.~\ref{Fig5} summarizes the simulation results across these scenarios. As shown in Fig.~\ref{Fig5}(a), the worst BER is $8.71 \times 10^{-2}$, and converges to zero when the SINR exceeds \qty{14}{dB} for all cases. This supports the argument that an SINR higher than \qty{13}{dB} can generally be regarded as sufficient for establishing a reliable communication link. Along with BER, Fig.~\ref{Fig5}(b) shows the error vector magnitude (EVM) performance as a function of SINR. The EVM quantifies the distortion between the received and ideal constellation points, which can be calculated as 
\begin{align}\label{Eq11}
	\mathrm{EVM} =  \sqrt{ \left( \frac{\sum_{n=1}^{N} \sum_{m=1}^{M} \left| s_{n,m} - \hat{s}_{n,m} \right|^2}{\sum_{n=1}^{N} \sum_{m=1}^{M} \left| s_{n,m} \right|^2} \right)}.
\end{align}
$s_{n,m}$ denotes the actual transmitted symbol and $\hat{s}_{n,m}$ denotes the corresponding received symbol at the $n$th subcarrier and $m$th symbol. A lower EVM value indicates better demodulation accuracy and, consequently, higher communication reliability. As the SINR increases, EVM performance improves consistently across all scenarios. Notably, increasing the bistatic transmit power improves EVM, since the resulting stronger bistatic LoS path enhances the robustness of communication link against both noise and monostatic interference. In addition, for a fixed transmit power, a higher monostatic RCS tends to result in slightly degraded EVM performance, as the stronger monostatic reflection introduces more interference to the bistatic communication performance. Lastly, Fig.~\ref{Fig5}(c) illustrates the monostatic radar image SINR performance as a function of SINR for different transmit powers and RCS conditions. The monostatic image SINR exhibits a nearly linear degradation with increased SINR. It is also observed that the influence of transmit power $P_{\text{Tx}}$ is negligible in this case, indicating that the radar sensing performance is mainly determined by the power ratio between the monostatic and bistatic components. 

In this work, we did not consider the case where SINR is below \qty{0}{dB}. This is because as shown in the case where $\mathrm{SINR} =$ \qty{0}{dB}, the BER is nearly $0.1$ which is problematic in practical scenarios. Also, such a high error rate severely degrades the sensing performance, where communication accuracy directly affects radar processing. Moreover, despite the fact that S\&C for synchronization benefits from correlation gain, false detection or missed detection probability of the starting point is highly expected when SINR is below 0 dB.

\section{Proposed SIC Method}\label{Sec3}
In this section, we present the proposed SIC method for multistatic ISAC systems. Core principle of the SIC-based approach is to determine whether to cancel the monostatic reflection or the bistatic LoS communication signal to increase the performance of both sensing and communications. Therefore, we first explain our approach of determining at which point to cancel out monostatic or bistatic LoS. Then we introduce our proposed cancellation method for each monostatic reflection and bistatic LoS.

\subsection{Prerequisite for Cancellation}\label{Sec3_1}
As mentioned in the previous paragraph, the core principle of the proposed SIC-based approach is to adaptively cancel either the monostatic reflection or the bistatic LoS signal, depending on their relative power contributions to the received signal. To enable this decision, the SIC process begins with removing the direct coupling component, which is the strongest interferer in multistatic ISAC systems. The cancellation is implemented using the same procedure employed for monostatic reflection cancellation, described in the following subsection. Once the direct coupling has been successfully canceled, the receiver must determine whether to cancel the monostatic reflection or the bistatic LoS component. In principle, this decision should be based on the SINR measured at the receiver. However, accurately estimating the SINR in practical scenarios is highly challenging, as the receiver lacks access to reference signals corresponding to isolated monostatic or bistatic path. In addition, the lack of precise knowledge about the transmitted symbols, channel responses, and path-specific attenuation further complicates the estimation, making direct SINR measurement infeasible. 

To address this limitation, we adopt the monostatic radar image SINR as a decision metric. When the SINR is low, synchronization algorithms such as those used to detect the signal start point can fail, resulting in missed detections or incorrect alignment of bistatic-oriented signals. Such errors hinder accurate channel estimation and make any cancellation unreliable or even infeasible. In contrast, monostatic radar processing remains functional under low SINR, even when the target is partially or completely masked by noise prior to signal processing. In addition, the radar image SINR can directly reflect the strength of monostatic reflection, making it a reliable indicator of which component dominates the received signal. The image SINR is computed by first identifying the strongest peak value in the normalized range-Doppler map, and then subtracting the average power measured over a predefined noise region. In multi-target scenarios, the SINR can be evaluated based on the strongest target response.

Based on the estimated monostatic image SINR, we employ a threshold-based decision strategy to select the appropriate component for cancellation. If the monostatic image SINR exceeds a predefined threshold, it indicates that the monostatic reflection is strong enough to interfere with the bistatic signals. In such case, cancelling the monostatic reflection may enhance the communication and bistatic imaging performance. On the other hand, if the monostatic image SINR is below the threshold, it implies that the bistatic LoS signal is stronger, which can mask the monostatic reflection and degrade sensing performance. To increase the monostatic sensing performance in such case, the bistatic LoS component is canceled. In this work, we set the threshold of monostatic image SINR as 45 dB. Before describing the cancallation procedures in detail, we first outline the following three key considerations that would provide important context for the proposed SIC method:
\begin{itemize}
	\item First, as mentioned earlier, direct coupling is canceled in digital domain using the proposed monostatic reflection cancellation method that will be explained in next subsection. This is possible because the direct coupling can be regarded as virtual target that is located at zero range with zero Doppler frequency. 
	
	\item Second, following the cancellation of direct coupling, the system determines whether to cancel out the monostatic reflection or the bistatic LoS path. However, the bistatic reflection is not considered as a candidate which needs to be canceled out. This is because the bistatic LoS signal is generally stronger than the bistatic reflection. While it is true that bistatic reflection also acts as interference from the perspective of monostatic sensing, their relative impact is much smaller than that of the bistatic LoS signal. Furthermore, except for rare situations where the bistatic reflection is nearly as strong as the LoS path such as when the target is located near the bistatic Tx or Rx, the processing gain is sufficient to tolerate such interference. 
	
	\item Third, the choice of \qty{45}{dB} as the threshold is not based on a strict mathematical derivation, but on heuristic observations. There are two main reasons for this selection. First, as shown in Figs.~\ref{Fig4}(a) and (c), SINR above \qty{14}{dB} (i.e., corresponding to a monostatic image SINR of \qty{52.57}{dB}) is sufficient to ensure error-free communication. Thus, cancelling the monostatic reflection beyond this point offers additional benefit only for sensing, not for communcations. However, this analysis only accounts for thermal noise, whereas in practice, additional hardware impairments can further degrade the SINR. To account for this discrepancy and ensure that the system operates reliably even under real-world conditions, we conservatively adjust the SINR threshold to approximately \qty{17}{dB} (i.e., corresponding to a monostatic image SINR of \qty{49.56}{dB}). Second, from the sensing perspective, a minimum image SINR of 15 to \qty{20}{dB} is required for target detections \cite{TI}. Along with minimum required image SINR, prior studies have shown that the peak power in the radar image can differ by 20–\qty{25}{dB} between high- and low-RCS targets \cite{Pedestrian_Trucks}. As a result, practical systems may demand up to \qty{40}{dB} of image SINR to ensure robust detection across a wide range of target types. Taking both sensing and communication requirements into account, we select a mean value, which is \qty{45}{dB}, as a reasonable threshold for our SIC framework.
\end{itemize}

\subsection{Cancellation Method for Monostatic Reflection}\label{Sec3_2}

As shown in Fig. \ref{Fig4}(a), even after the cancellation of direct monostatic coupling, there are scenarios in which the power of the monostatic reflection remains comparable to, or only slightly below, that of the bistatic LoS signal. In such cases, the communication link is no longer reliable due to the insufficient dominance of the LoS communication path. It is therefore necessary to cancel out the monostatic reflection component to enable stable communication peformance. It is important to note that monostatic system has several advantages over the bistatic system. Specifically, the receiver has prior knowledge of the starting point of the monostatic signal, as well as the transmitted symbols. This allows us to skip additional processing steps such as channel estimation or equalization for demodulating the received signal, which would otherwise be required in bistatic scenarios. 

Let us assume that the received signal $y(t)$ passes through a LNA, BPF, and a down conversion stage before being digitized by an ADCs. The resulting $\mathrm{I}$ and $\mathrm{Q}$ samples are combined to form complex baseband samples. From these digitized samples, a total of $(N_{\text{cp}}+N) \times M$ samples are collected starting from the known transmission time, and perform S/P conversion to obtain one OFDM frame $\mathbf{Y}_{\text{mono}} \in \mathbb{C}^{(N_{\text{cp}} + N) \times M}$. After discarding the CP, we transform the signal into the frequency-domain using DFT. Since received samples contain both the channel response and the transmitted symbols, we remove the known transmit symbols by element-wise division, resulting in the estimated radar channel response $\hat{\mathbf{H}}_{\text{mono}} \in \mathbb{C}^{N \times M}$. Finally, we apply an IDFT and DFT along the subcarrier and symbol axes to generate a 2D range-Doppler (RD) map . We denote the estimated range and Doppler frequency as $\hat{R}_{\text{mono}}$ and $\hat{f}_{\text{D,mono}}$, respectively, where $\hat{R}_{\text{mono}} = (c \cdot \hat{\tau}_{\text{mono}})/2$ with $\hat{\tau}_{\text{mono}}$ being the estimated propagation delay. 

When the image SINR of the monostatic RD map exceeds \qty{45}{dB}, the time-domain received signal is reconstructed based on the estimated target parameters, including the range, Doppler frequency, and peak amplitude from the RD map. Once the target delay and Doppler are estimated, the radar channel response is generated as
\begin{align}\label{Eq12}
	\hat{h}^{n,m}_{\text{mono,recon}} =  \text{e}^{-\text{j} 2\pi n \hat{\tau}_{\text{mono}} \Delta f} \text{e}^{\text{j} 2\pi m T_{\text{sym}} \hat{f}_{\text{D,mono}}},
\end{align}
where $\Delta f$ and $T_{\text{sym}}$ represent the subcarrier spacing and symbol duration including CP, respectively. By applying an IDFT and DFT along the subcarrier and symbol axes to $\hat{\mathbf{H}}_{\text{mono,recon}}$, the same range and Doppler indices as those in the received RD map can be reproduced. To ensure that the reconstructed response matches the power level of the actual signal, amplitude normalization is performed using the peak value of the original RD map at the target bin. This approach inherently accounts for frequency-dependent gain variations due to RF impairments (e.g., DAC/ADC ripple or front-end filtering), as the target’s observed peak already reflects the system transfer characteristics. The normalization factor is computed as the ratio of the RD map peak to the reconstructed channel amplitude $\alpha_{\text{recon}}$, and the final radar channel is scaled accordingly.
\begin{align}\label{Eq13}
	\hat{\mathbf{H}}^{'}_{\text{mono,recon}} = \left(\frac{\alpha_{\text{received}}}{\alpha_{\text{recon}}}\right) \times \hat{\mathbf{H}}_{\text{mono,recon}}. 
\end{align}
%
Following this normalization, we regenerate the time-domain monostatic reflection by multiplying the normalized channel response with the known transmit symbols, prepending CPs, and performing P/S conversion. The resulting signal is then subtracted from the original received time-domain signal to cancel out the monostatic reflection, thereby enhancing bistatic communication and sensing performances. As discussed previously, the same procedure is also employed for canceling direct coupling.

\subsection{Cancellation Method for Bistatic LoS Path}\label{Sec3_3}
On the other hand, there are scenarios where the power of the bistatic LoS path becomes significantly higher than that of the monostatic reflections, as shown in Fig. \ref{Fig4}(c). In such case, the SINR of the monostatic radar image falls below the predefined cancellation threshold, indicating that the dominant component in the received signal is the bistatic LoS path. While this strong LoS component is favorable for communications, it can severely degrade the monostatic radar image and, in particular, mask the presence of weak targets. Unlike the monostatic cancellation scenario, where the interfering signal has known structure due to the fact that it is derived from the own system, the LoS communication signal requires synchronization. Therefore, after having the down-converted time-domain samples, we first have to estimate the offsets to further process the received signal correctly.

The full synchronization process begins by estimating a coarse STO and fractional CFO using S\&C applied to the first preamble symbol. These estimates are then used to extract the second preamble symbol, which enables refinement of the integer CFO, yielding a complete CFO estimate. With this full CFO, frequency offset correction is applied to the first preamble region of the received signal. Next, a fine STO estimate is obtained via correlation with the known transmit preamble, and the entire frame is corrected accordingly. The signal is then S/P converted, and CPs are removed. After transforming the resulting symbols into the frequency domain via DFT, pilot subcarriers are extracted and used to estimate the SFO using the TITO algorithm. The estimated SFO is then compensated through resampling of the received time-domain signal. Finally, since the signal is resampled, it undergoes a CP removal and DFT operation once more, after which residual STO and CFO are estimated based on pilot subcarriers. Detailed explanation of offset estimations for synchronization can be found in \cite{TITO}. Once the offset estimations are completed, the received signal is compensated for these offsets.
After synchronization, we collect a total of $(N_{\text{cp}}+N) \times M$ samples from the estimated starting point of symbol, as in the same manner as the monostatic cancellation procedure. 


%
\begin{figure*}[!t]
	\begin{center}
		\includegraphics[width = 1\linewidth]{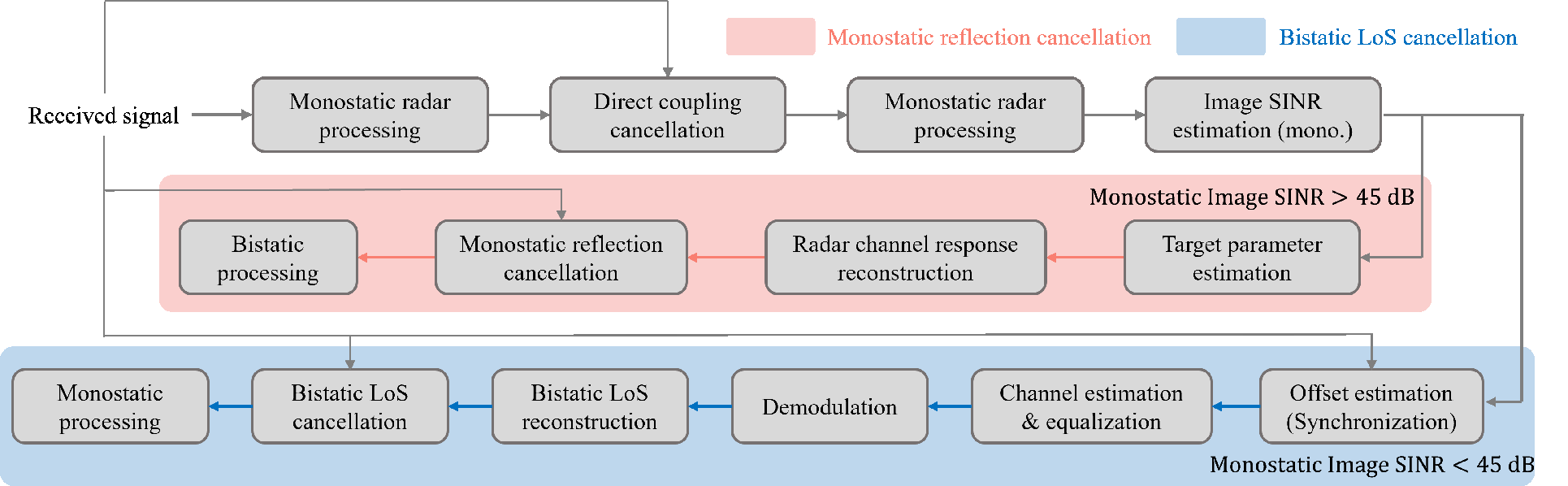} 
		\caption{Overall procedure of the proposed interference cancellation framework.}\label{Fig7}
	\end{center}
\end{figure*}

Next, channel estimation is performed using the extracted pilot symbols, and the results are interpolated to yield the full channel response over the entire OFDM frame, denoted by $\hat{\mathbf{H}}_{\text{bi}}$. The transmitted symbol matrix $\hat{\mathbf{X}}_{\text{bi}}$ is then obtained by equalizing the frequency-domain representation of the received signal, where each subcarrier is divided by its corresponding estimated channel coefficient. Using the estimated synchronization parameters, channel response, and transmit symbols, the time-domain bistatic LoS signal can be reconstructed in two steps. First, the estimated symbol matrix is element-wise multiplied with the estimated channel response (i.e., $\hat{\mathbf{H}}_{\text{bi}} \odot \hat{\mathbf{X}}_{\text{bi}}$), where $\odot$ denotes element-wise multiplication. The resulting frequency-domain signal is then transformed into the time domain via IDFT, followed by the addition of a cyclic prefix. Finally, the signal is transformed back to its original asynchronous form by reapplying the estimated timing and frequency offsets. The full procedure of the proposed interference cancellation method is summarized in Fig.~\ref{Fig7}.

Note that the proposed method is based on the SIC strategy, it allows to cancel out the interference sequentially also in the multi-target scenarios. Specifically, when multiple reflections corresponding to different targets are present in the received signal, the image SINR is computed individually for each peak in the RD map. Monostatic reflections with SINR values exceeding the predefined threshold are cancelled one by one, in descending order of peak intensity. After each cancellation, the RD map is updated and re-evaluated to check whether additional targets still exceed the threshold. This iterative process continues until no remaining target satisfies the cancellation condition. At this point, bistatic LoS cancellation is applied to suppress the dominant communication signal, enabling the recovery of weaker monostatic reflections that would otherwise be masked. On the other hand, if none of the targets exceed the SINR threshold from the beginning, it is assumed that the bistatic LoS path dominates. In such cases, bistatic LoS cancellation is performed first, allowing weak monostatic reflections to be detected without strong interference.

\subsection{Cancellation Result}\label{Sec3_4}

The performance evaluation of the proposed interference cancellation framework under two representative scenarios is presented in this section. In the first scenario (i.e., Case 1), the monostatic radar image SINR exceeds the cancellation threshold after direct coupling removal, with an estimated value of \qty{63.95}{dB}. The target ranges are estimated as \qty{12.06}{m} for the monostatic path and \qty{5.69}{m} for the bistatic path, with corresponding Doppler frequencies of \qty{0}{kHz} and \qty{-0.46}{kHz}, respectively. The CFO and SFO are estimated to be \qty{14.52}{kHz} and \qty{49.8372}{ppm}. In the second scenario (i.e., Case 2), the monostatic image SINR falls below the threshold, with an estimated value of \qty{37.45}{dB}. The monostatic and bistatic target ranges are estimated as \qty{15.58}{m} and \qty{9.89}{m}, respectively, and the corresponding Doppler frequencies are \qty{-1.91}{kHz} and \qty{-0.27}{kHz}. The estimated CFO and SFO for this case are \qty{0.267}{kHz} and \qty{30.12}{ppm}, respectively. All other simulation parameters are based on the settings defined in Table~\ref{Tab1}, while the corresponding ground truth values for each scenario are summarized in Table~\ref{Tab2}.

\begin{table}[!b]	
	\centering
	\caption{Parameter configurations for two representative cases}
	\label{Tab2}
	\renewcommand{\arraystretch}{1.3}  
	\setlength{\tabcolsep}{10pt}              
	\begin{tabular}{l||cc}
		\hline \hline
		\rowcolor{Silver}
		\textbf{Parameters} & \multicolumn{2}{c}{\textbf{Values}}                    \\ \hline \hline
		& \multicolumn{1}{c|}{\textbf{Case 1}}       & \textbf{Case 2}       \\ \hline
		$R_\text{LoS}$      & \multicolumn{2}{c}{\qty{20}{m}}                    \\ \hline
		$R_\text{mono}$     & \multicolumn{1}{l|}{\qty{12.04}{m}}   & \qty{15.56}{m}  \\ \hline
		$R_\text{bi}$       & \multicolumn{1}{l|}{\qty{5.64}{m}}   & \qty{9.76}{m}   \\ \hline
		$f_\text{D,mono}$   & \multicolumn{1}{l|}{\qty{-0.08}{kHz}} & \qty{-1.98}{kHz} \\ \hline
		$f_\text{D,bi}$  & \multicolumn{1}{l|}{\qty{-0.48}{kHz}} & \qty{-0.36}{kHz} \\ \hline
		CFO   & \multicolumn{1}{l|}{\qty{10}{kHz}}    & \qty{0}{kHz}    \\ \hline
		SFO   & \multicolumn{1}{l|}{\qty{50}{ppm}}    & \qty{30}{ppm}   \\ \hline
	\end{tabular}
\end{table}

\begin{figure}[!t]
	\begin{center}
		\includegraphics[width = 0.47\linewidth]{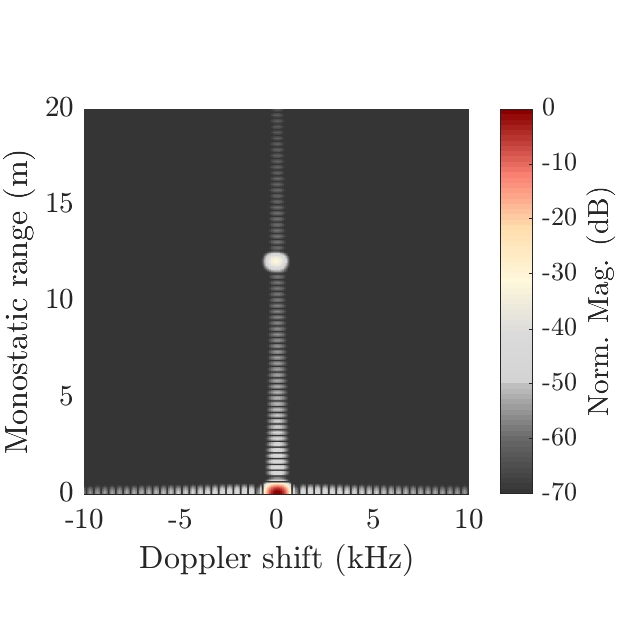} 
		\includegraphics[width = 0.47\linewidth]{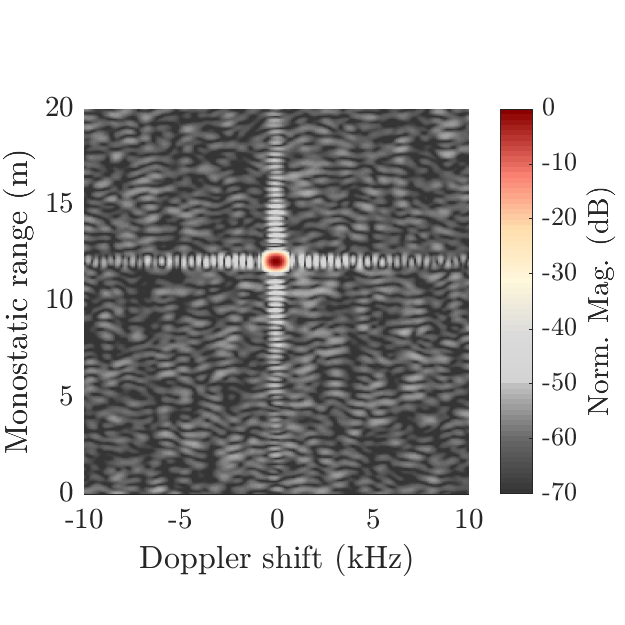}  
		\\ (a) \qquad\qquad\qquad\qquad\qquad (b) 
		\caption{Simulated monostatic RD map: (a) before direct coupling cancellation, (b) after direct coupling cancellation.}\label{Fig8}
	\end{center}
\end{figure}

\begin{figure}[!t]
	\begin{center}
		\includegraphics[width = 0.47\linewidth]{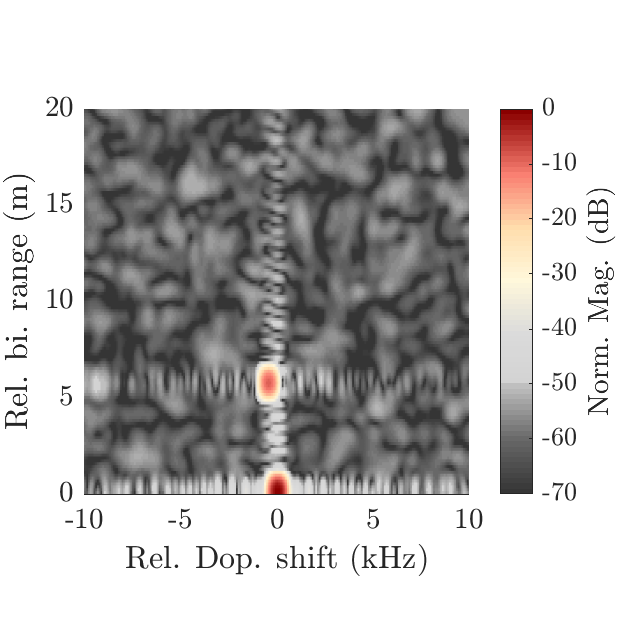}
		\includegraphics[width = 0.47\linewidth]{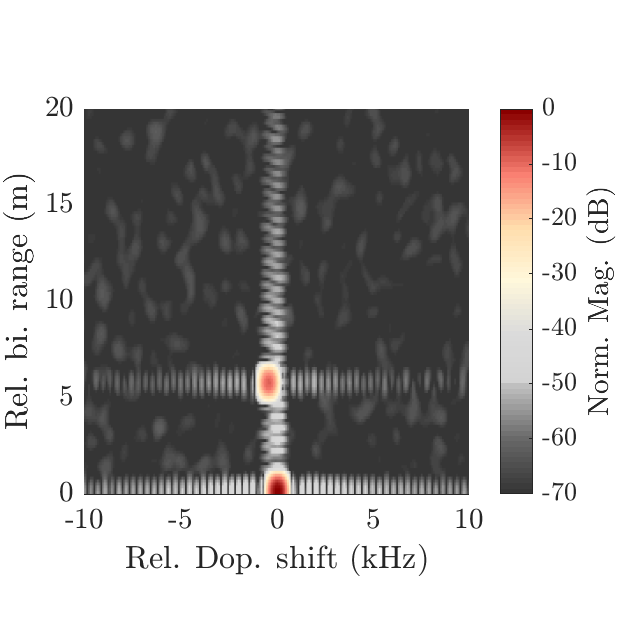} 
		\\ (a) \qquad\qquad\qquad\qquad\qquad (b) 
		\caption{Simulated bistatic RD map: (a) before monostatic reflection cancellation, (b) after monostatic reflection cancellation.}\label{Fig9}
	\end{center}
\end{figure}

\begin{figure}[!t]
	\begin{center}
		\includegraphics[width = 0.47\linewidth]{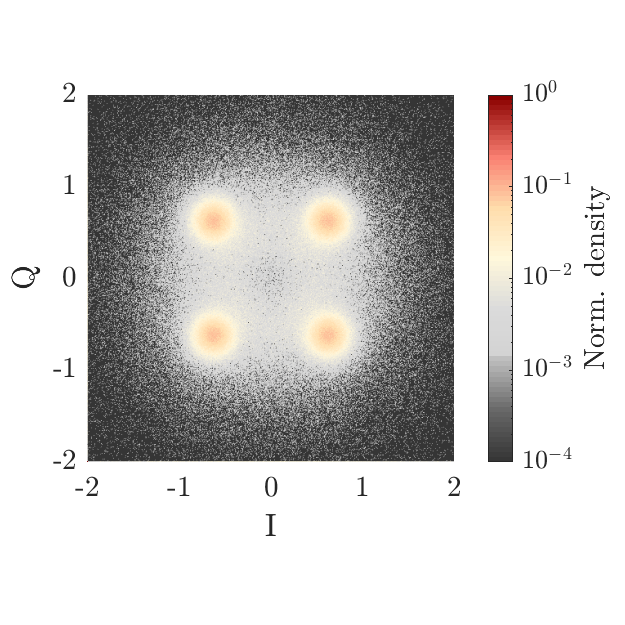}
		\includegraphics[width = 0.47\linewidth]{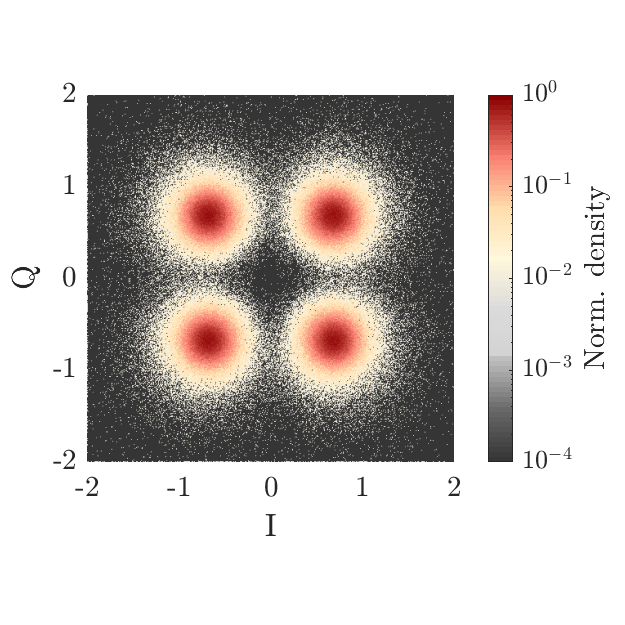} 
		\\ (a) \qquad\qquad\qquad\qquad\qquad (b) 
		\caption{Simulated constellation diagram after demodulation: (a) before monostatic reflection cancellation, (b) after monostatic reflection cancellation.}\label{Fig10}
	\end{center}
\end{figure}

As shown in Fig.~\ref{Fig8}(a), a strong peak appears at zero range and  zero Doppler, corresponding to the direct coupling component. Following the cancellation step described in Sec. \ref{Sec3_1}, we reconstruct the time-domain signal for coupling using the estimated range, Doppler frequency, and peak amplitude. After reconstruction, we subtract the reconstructed time-domain signal from originally received time-domain signal. Fig.~\ref{Fig8}(b) illustrates the resulting monostatic RD map after direct coupling cancellation. After removing the direct coupling component, the monostatic radar image SINR is re-calculated, and estimated to be \qty{63.95}{dB}, which exceeds the threshold. Based on the proposed SIC method, we then cancel the monostatic reflection to enhance the bistatic communication and sensing performances. Fig.~\ref{Fig9} presents the bistatic radar image before and after monostatic cancellation. As shown in the figure, the background noise level across the radar image is significantly reduced after cancellation. The bistatic image SINR improves from \qty{47.22}{dB} to \qty{54.03}{dB}. Furthermore, the constellation diagram shown in Fig.~\ref{Fig10} illustrates the improvement in communication performance. Specifically, the BER is reduced from $6.47 \times 10^{-2}$ to $5.91 \times 10^{-4}$, and the EVM is also reduced from \qty{-10.34}{dB} to \qty{-15.62}{dB}.

\begin{figure}[!t]
	\begin{center}
		\includegraphics[width = 0.47\linewidth]{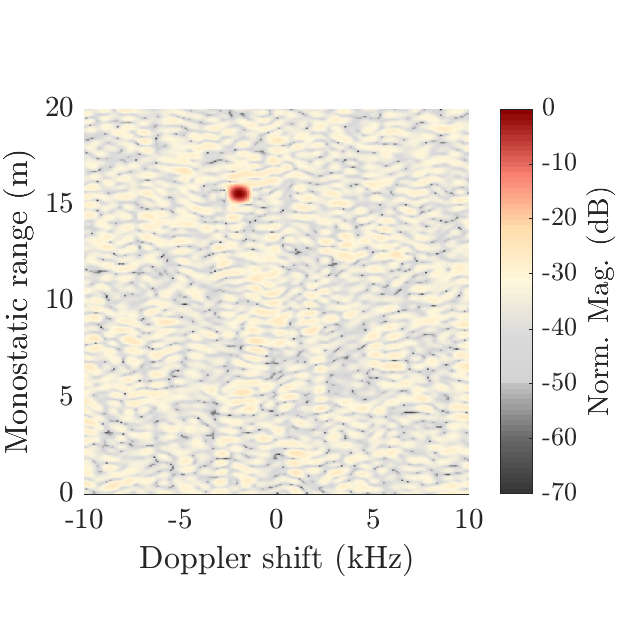}
		\includegraphics[width = 0.47\linewidth]{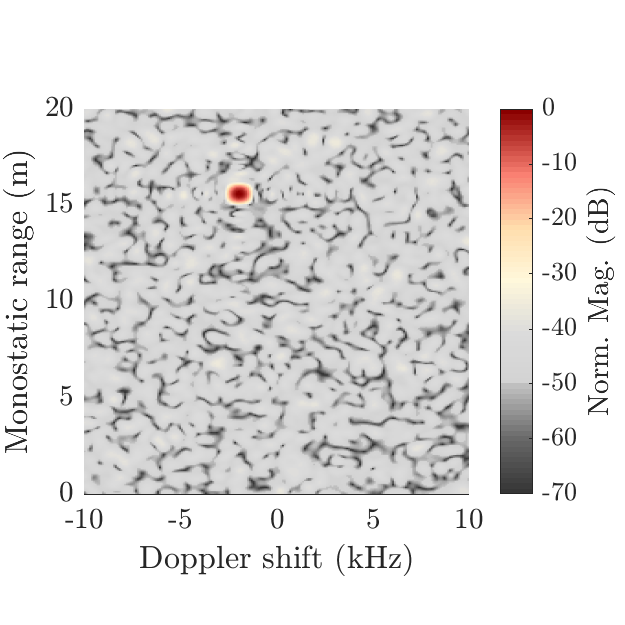} 
		\\ (a) \qquad\qquad\qquad\qquad\qquad (b) 
		\caption{Simulated monostatic RD map: (a) before bistatic LoS cancellation, (b) after bistatic LoS cancellation.}\label{Fig11}
	\end{center}
\end{figure}

The second scenario considers the case in which the monostatic radar image SINR falls below the threshold after direct coupling cancellation. This indicates that the power of the bistatic signal is relatively high compared to the monostatic reflection. In such cases, communication performance is expected to remain reliable without additional processing. However, strong interference from the bistatic signal, particularly the LoS component, can degrade the monostatic sensing performance. The result of canceling this interference is shown in Fig.~\ref{Fig11}, which illustrates the effect of bistatic LoS cancellation on the monostatic radar image SINR. Specifically, the monostatic image SINR increases from \qty{37.45}{dB} to \qty{47.67}{dB}.

\begin{figure*}[!t]
	\begin{center}
		\includegraphics[width = 0.24\linewidth]{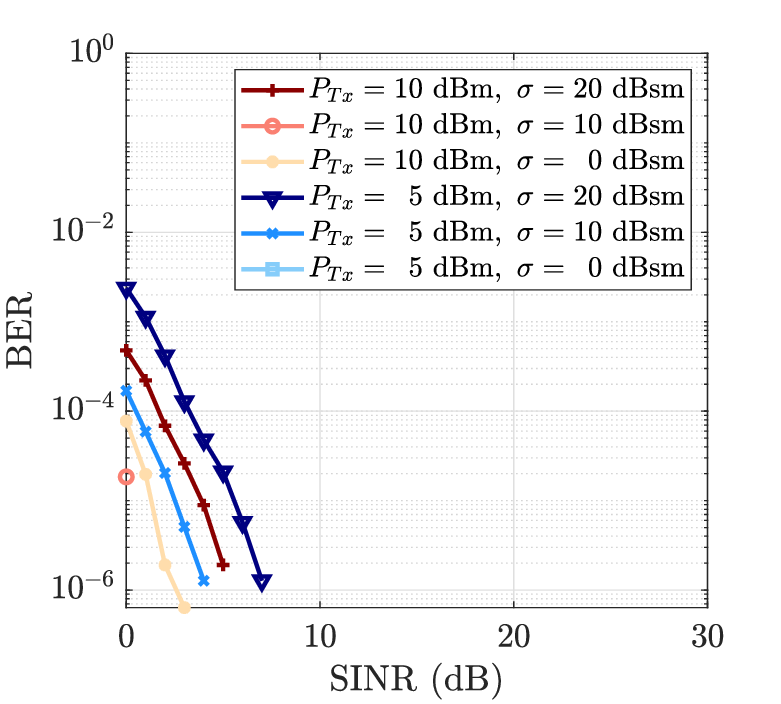}
		\includegraphics[width = 0.24\linewidth]{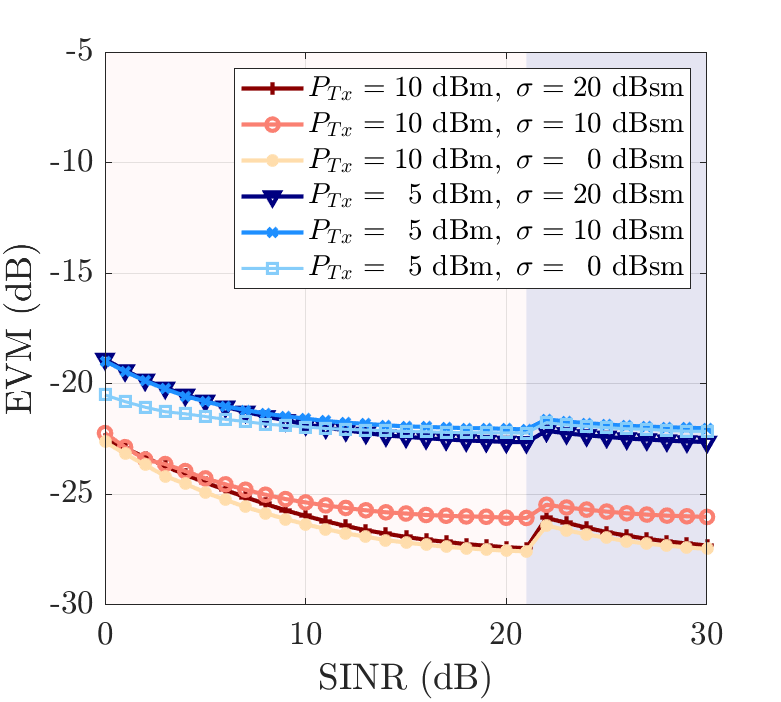} 
		\includegraphics[width = 0.24\linewidth]{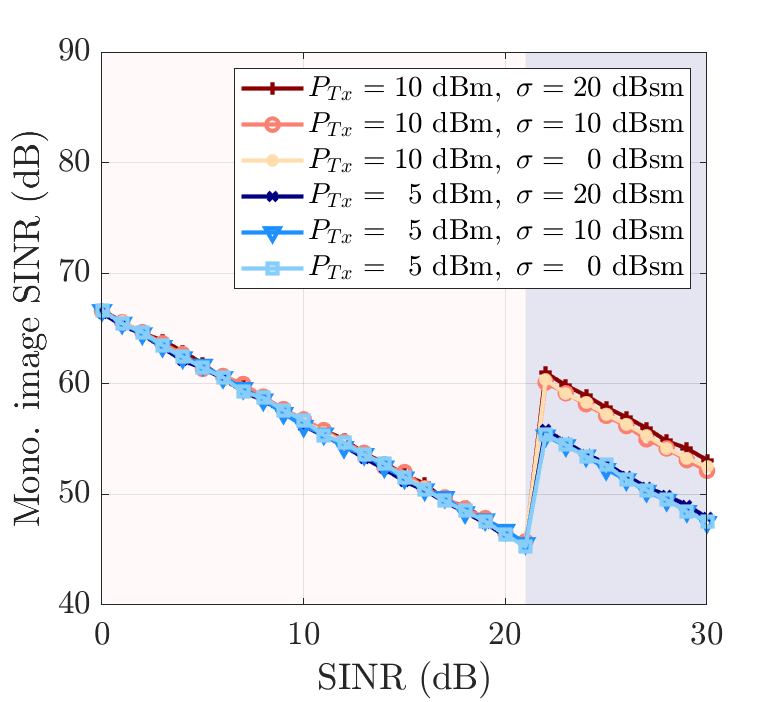}
		\includegraphics[width = 0.24\linewidth]{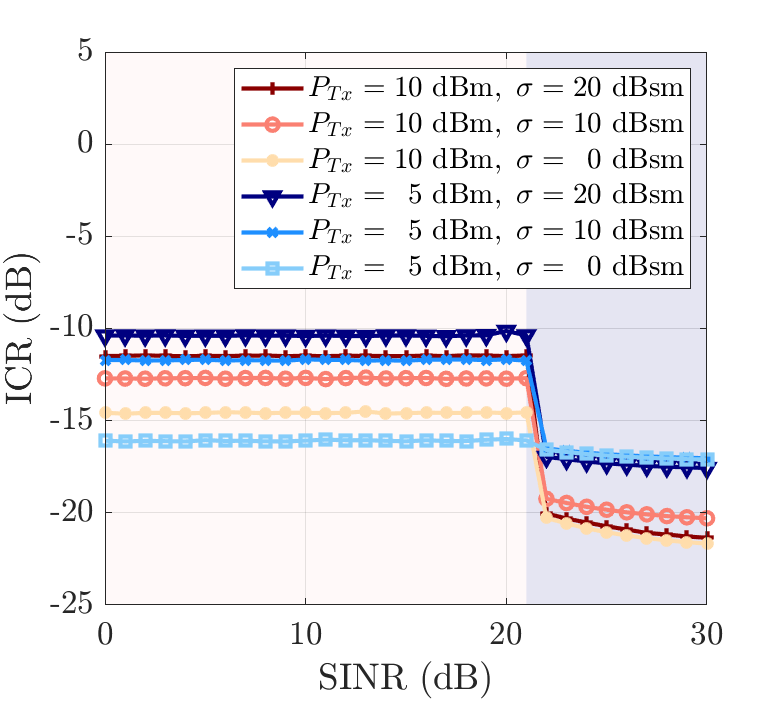} 
		\\ (a) \qquad\qquad\qquad\qquad\qquad\qquad (b) \qquad\qquad\qquad\qquad\qquad\qquad (c) \qquad\qquad\qquad\qquad\qquad\qquad (d)\\
		\caption{
			Results of sensing and communication performances in interference-cancelled scenarios with varied SINR values:
			(b) BER, (c) monostatic image SINR, (d) ICR. 
			The pink-shaded (\textcolor[rgb]{1.000,0.8906,0.8789}{\rule{0.7em}{1.5ex}}) region indicates the SINR range where monostatic reflection cancellation is applied, 
			while the blue-shaded (\textcolor[rgb]{0.0,0.0,0.5}{\rule{0.7em}{1.5ex}}) region marks the range where bistatic LoS cancellation is applied.
		}
		\label{Fig12}
	\end{center}
\end{figure*}

Followed by the cancellation results of the two representative scenarios, we further evaluate the effectiveness of the proposed SIC method using the same scenarios employed for interference analysis in Sec.~\ref{Sec2_3}. To quantify the cancellation performance, we define a metric called the interference cancellation ratio (ICR), which measures the relative reduction in interference power after cancellation. The ICR is defined as the logarithmic ratio between the power of the original reference signal and the residual interference power after subtracting the reconstructed interference component, as given by:
\begin{align}\label{Eq15}
	\mathrm{ICR} =  \sqrt{\left( \frac{ \int |y_{\text{ref}}(t) - y_{\text{recon}}(t)|^2 \, dt }{ \int |y_{\text{ref}}(t)|^2 \, dt } \right)}.
\end{align}
In \eqref{Eq15}, $y_{\text{ref}}(t)$ denotes the reference signal (i.e., the isolated path) originating from either the monostatic or bistatic transmitter, while $y_{\text{recon}}(t)$ represents the reconstructed interference component. The operator $|\cdot|$ indicates the $\ell_2$-norm. Compared to the baseline results in Fig.~\ref{Fig5}, the BER is reduced across all cases, with the worst-case value decreasing from 0.1 to $2.3 \times 10^{-3}$. Furthermore, whereas the BER in the baseline scenario required \qty{15}{dB} SINR to reach 0, the proposed cancellation method achieves BER $=0$ across all evaluated cases at \qty{7}{dB} SINR. Fig.~\ref{Fig12}(b) and \ref{Fig12}(c) show the corresponding EVM and monostatic radar image SINR, respectively. In both figures, the pink-shaded region indicates the range of SINR where monostatic reflection cancellation is applied, while the blue-shaded region marks the range where bistatic LoS cancellation is performed. Compared to the baseline results in Fig.~\ref{Fig5}, the EVM shows a clear improvement. The average EVM reduction is \qty{12.39}{dB} when the transmit power is set to \qty{10}{dBm}, and \qty{7.97}{dB} when it is set to \qty{5}{dBm}. For the monostatic radar image SINR, a sharp increase is observed at the decision threshold where the cancellation method switches from monostatic reflection suppression to bistatic LoS cancellation. After applying bistatic LoS cancellation, the average image SINR gain is \qty{13.5}{dB} for the \qty{10}{dBm} transmit power case, and \qty{9.71}{dB} for the \qty{5}{dBm} case.

\begin{figure*}[!t]
	\begin{center}
		\includegraphics[width = 0.32\linewidth]{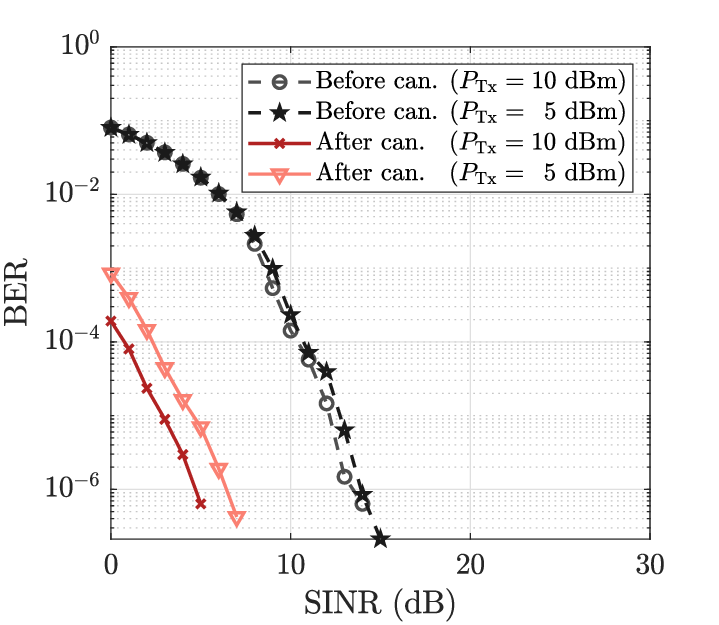}
		\includegraphics[width = 0.32\linewidth]{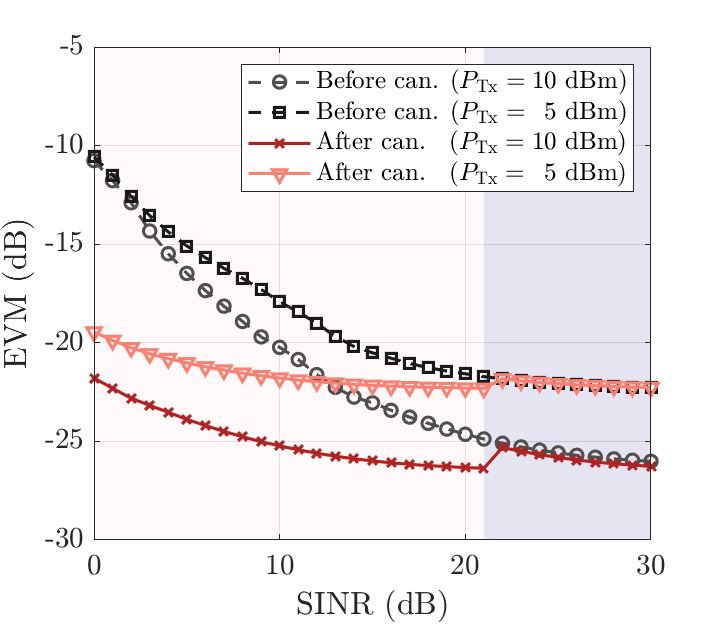} 
		\includegraphics[width = 0.32\linewidth]{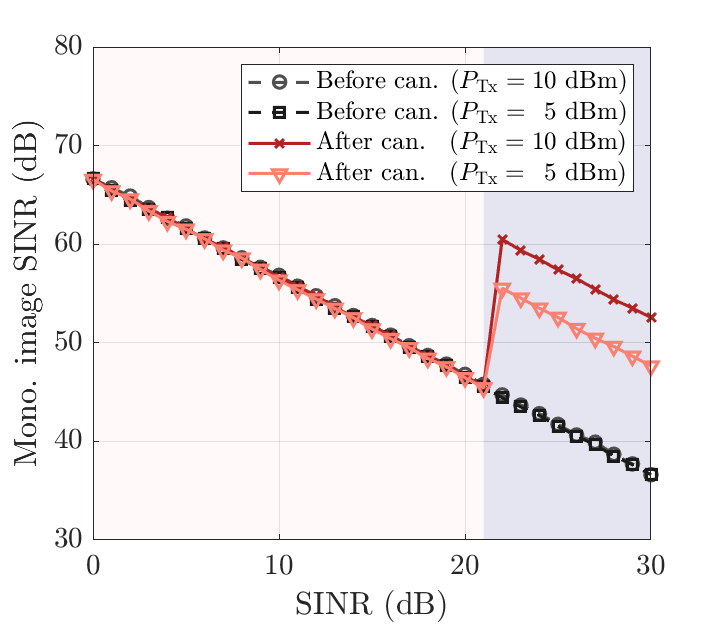}
		\\ (a) \qquad\qquad\qquad\qquad\qquad\qquad (b) \qquad\qquad\qquad\qquad\qquad\qquad (c) \\
		\caption{
			Comparison of sensing and communication performance metrics before and after the interference cancellation across different SINR values in the simulation: (a) BER, (b) EVM, and (c) monostatic image SINR.}
		\label{Fig13}
	\end{center}
\end{figure*}

Finally, Fig.~\ref{Fig12}(d) illustrates the ICR across various SINR values. A notable observation in the monostatic cancellation region is that the reconstruction performance improves as the RCS of target decreases when the transmit power is fixed. This is because weaker monostatic reflections result in lower absolute signal power, thereby reducing the impact of any reconstruction mismatch in absolute terms. In contrast, when the RCS is large, even small errors in amplitude or phase estimation can lead to larger absolute residuals after cancellation, which degrades the performance. Therefore, the residual power becomes more pronounced at high RCS values despite similar relative reconstruction accuracy, creating the illusion of worse cancellation performance. For bistatic LoS cancellation, the ICR remains largely unaffected by the target’s RCS, as the LoS component does not depend on target reflection characteristics. Instead, the cancellation performance in this region is more strongly influenced by the transmit power. Higher transmit power leads to stronger LoS signals that are less susceptible to noise and channel distortion, enabling more accurate reconstruction. Moreover, when directly comparing the two types of cancellation, bistatic LoS cancellation consistently yields better performance across all scenarios. This can be mainly attributed to the deterministic and stable nature of the bistatic LoS path. Unlike monostatic reflections, which depend on target-specific characteristics such as RCS, the bistatic LoS signal propagates through a direct path that is less affected by environmental variations. As a result, its channel response is simpler and more predictable, enabling more accurate reconstruction. In addition to the results shown in Fig.~\ref{Fig12}, Fig.~\ref{Fig13} summarizes the overall performance changes before and after interference cancellation across varying the SINR conditions. The plots show the trends in BER, EVM, and monostatic image SINR, clearly showing that the proposed method improves both sensing and communication performance by effectively canceling interference.

\subsection{Compuational Complexity Derivation for the Proposed Cancellation Method}\label{Sec3_5}

The computational complexity of the proposed SIC method is determined by the signal reconstruction procedures for monostatic and bistatic interference components. For both cases, the complexity is expressed in terms of $N$ and $M$. For monostatic reflection cancellation, the dominant operations include frequency-domain transformation, radar channel estimation, and RD map processing. Specifically, after performing the DFT on each of the $M$ OFDM symbols, which requires a complexity of $\mathcal{O}(MN\log N)$, the channel response is estimated by element-wise division with known transmit symbols, requiring $\mathcal{O}(MN)$. The resulting channel matrix is transformed into the RD map using an IDFT/FFT, yielding an additional complexity of $\mathcal{O}(MN\log N + MN\log M)$. Channel reconstruction, which uses the estimated range, Doppler, and peak values, involves evaluating complex exponentials over the full $N \times M$ grid, with complexity $\mathcal{O}(MN)$. This is followed by normalization, IDFT, and cyclic prefix insertion, all of which remain within $\mathcal{O}(MN\log N)$. Therefore, the overall computational complexity for monostatic cancellation is in the order of $\mathcal{O}(MN\log N + MN\log M)$.

In the case of bistatic LoS cancellation, the key steps involve estimating the unknown transmit symbols by dividing the frequency domain received signal by the estimated channel response, with a complexity of $\mathcal{O}(MN)$. This is followed by signal reconstruction through element-wise multiplication of the estimated symbols and channel, and an IDFT to return to the time domain. Similar to the monostatic case, these steps require a complexity of $\mathcal{O}(MN\log N)$. Additionally, the bistatic cancellation process includes resampling to compensate for the estimated SFO, which introduces further computational cost. This resampling is necessary because the sampling time offsets caused by SFO result in non-uniform time grids, requiring interpolation to reconstruct the signal on a uniform grid. This step typically incurs a complexity in the order of $\mathcal{O}(MN)$ to $\mathcal{O}(MN\log N)$, and should be taken into account in scenarios with non-negligible SFO. Therefore, the overall proposed SIC method remains computationally efficient, relying primarily on DFT-based operations and simple element-wise computations.

\section{Proof-of-concept Measurement}\label{Sec4}
\begin{figure}[!t]
	\begin{center}
		\includegraphics[width = 0.6\linewidth]{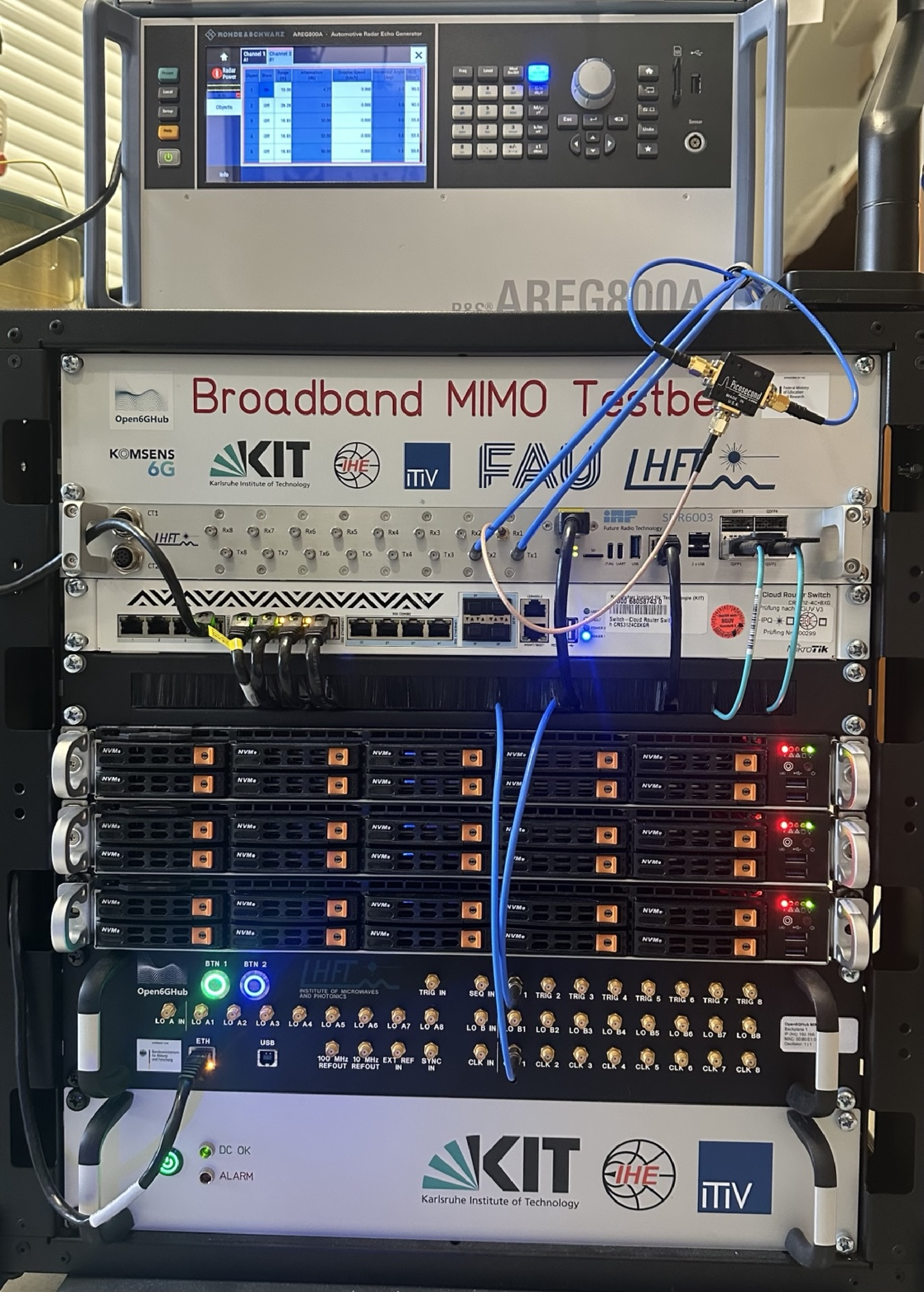}
		\caption{Measurement setup with ISAC testbed and radar target simulator and its simplified block diagram.}\label{Fig14}
	\end{center}
\end{figure}

Proof-of-concept measurements were conducted to validate the proposed SIC method in practical settings. The overall measurement setup is illustrated in Fig.~\ref{Fig14}. This setup uses an ISAC testbed and the AREG800A, a radar target generator from Rohde \& Schwarz for emulating target reflections. Additionally, a power combiner was used to merge both mono- and bistatic signals prior to feeding them into the receiver chain. Further details regarding the ISAC testbed can be found in \cite{Testbed1} and \cite{Testbed2}.

It is worth mentioning that the measurement setup is characterized by four important characteristics. First, a single BB module supporting up to 8 Tx and 8 Rx channels was used. As a result, both mono- and bistatic signals were generated within the same BB module, eliminating the need for synchronization between bistatic transmitters and receivers. Second, the measurements were conducted at a carrier frequency of \qty{3.68}{GHz} with a bandwidth of \qty{0.98}{GHz}. Although this frequency was not covered in the simulation results shown in \ref{Sec2_3} and \ref{Sec3_3}, the proposed SIC method still demonstrated robust performance, confirming its applicability across a broad frequency range. Third, direct coupling was not considered due to measurement constraints; hence, direct coupling cancellation is assumed to be completed during the interference cancellation process. Lastly, extracting reference paths for both mono- and bistatic components were highly challenging, which makes direct ICR calculation infeasible. Nevertheless, SINR could be estimated based on the attenuation characteristics provided from the AREG800A.

Similar to the simulations, we first show the results under two representative cases: one where the monostatic radar image SINR exceeds the predefined threshold, and another where it falls below. Then, we show the measurement results where we vary the SINR values from 0 dB to 30 dB. Apart from the carrier frequency, bandwidth, and pilot type, all system parameters used in the measurements were identical to those employed in the simulations. As mentioned earlier, the carrier frequency and bandwidth were set to \qty{3.68}{GHz} and \qty{0.98}{GHz}, respectively. Furthermore, while a lattice-type pilot pattern was used in simulation, a block-type pilot with an interval of every two OFDM symbols was adopted for the measurement. In addition, the monostatic range and Doppler were set to \qty{20}{m} and \qty{0}{Hz}, while the bistatic relative range, defined as $(R_{\text{Tx}\rightarrow \text{Tgt}} + R_{\text{Tgt}\rightarrow \text{Rx}} - R_{\text{LoS}})$, and relative bistatic Doppler were set to \qty{10}{m} and \qty{0}{Hz}. 

\begin{figure}[!t]
	\begin{center}
		\includegraphics[width = 0.48\linewidth]{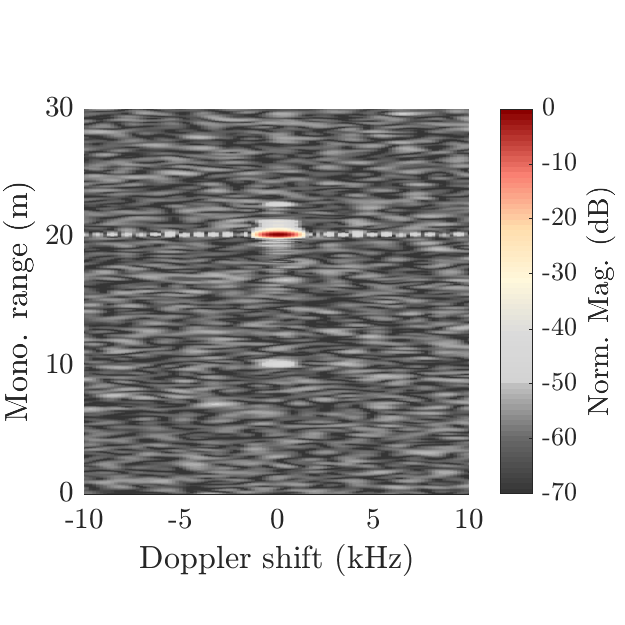}
		\includegraphics[width = 0.48\linewidth]{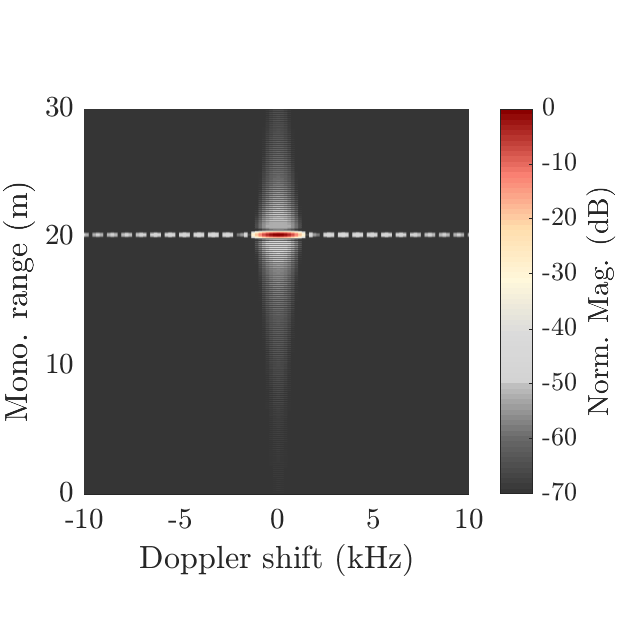}
		\\ (a) \qquad\qquad\qquad\qquad\qquad (b) 
		\caption{Measured monostatic RD maps: (a) Original RD map obtained from received signal, (b) reconstructed RD map with target estimates.}\label{Fig15}
	\end{center}
\end{figure}

\begin{figure}[!t]
	\begin{center}
		\includegraphics[width = 0.48\linewidth]{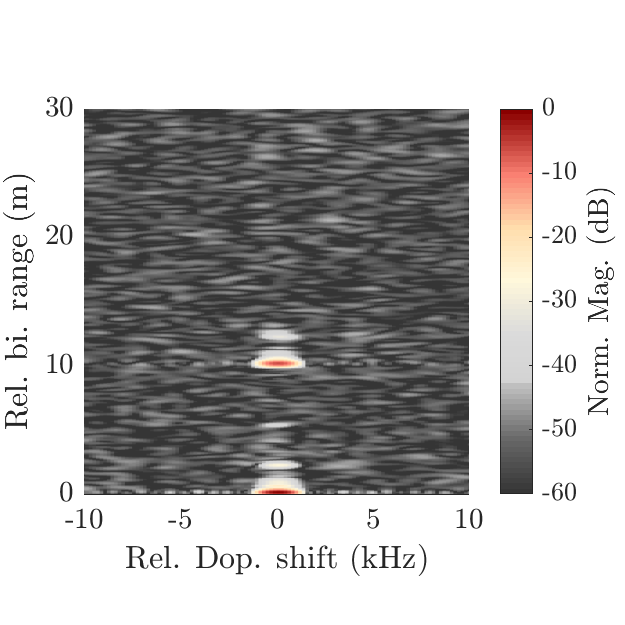}
		\includegraphics[width = 0.48\linewidth]{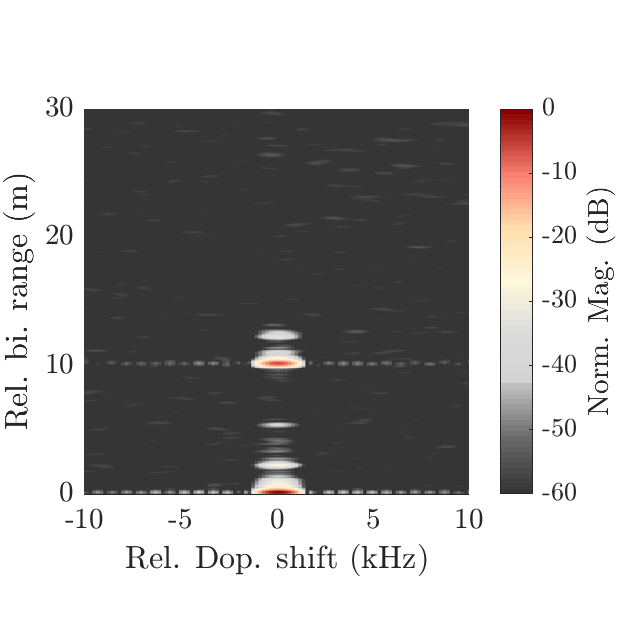}
		\\ (a) \qquad\qquad\qquad\qquad\qquad (b) 
		\caption{Measured bistatic RD maps: (a) before monostatic relfection cancellation, (b) after monostatic reflection cancellation.}\label{Fig16}
	\end{center}
\end{figure}

\begin{figure}[!t]
	\begin{center}
		\includegraphics[width = 1\linewidth]{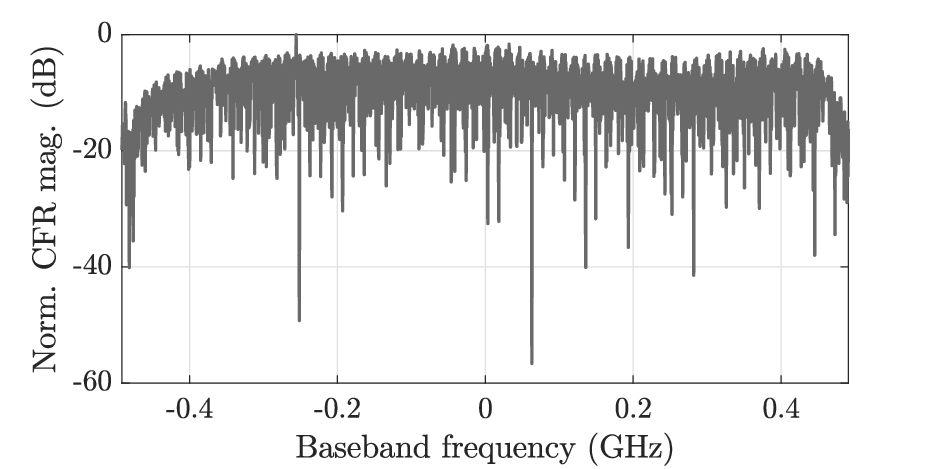}
		\caption{Estimated normalized channel frequency response at the baseband frequency.}\label{Fig17}
	\end{center}
\end{figure}

Starting from the extraction of valid samples corresponding to monostatic reflections, we generate the monostatic RD map shown in Fig.~\ref{Fig15}(a). In this scenario, the estimated SINR was \qty{2}{dB}. The resulting monostatic RD map yields an image SINR of \qty{61.91}{dB}. Since this exceeds the predefined threshold, monostatic reflection cancellation is applied based on the estimated target parameters. The reconstructed radar channel is presented in Fig.~\ref{Fig15}(b). In parallel, Fig.~\ref{Fig16} shows the bistatic RD maps before and after monostatic cancellation. As shown in bistatic radar image without cancellation, noise artifacts are observed, while those surrounding noise are removed after interference cancellation. It is important to note that the target-like peaks appearing near \qty{2.1}{m}, \qty{5.3}{m}, and \qty{12.4}{m} in the bistatic RD maps are attributed to the non-flat frequency response of the measured channel as shown in Fig.~\ref{Fig17}. This non-ideal response arises from hardware impairments, such as cable reflections, and internal reflections in the power combiner. However, these artifacts are much weaker than the target peak, and therefore can be neglected. Moreover, because the monostatic reflection affects the processing of the bistatic LoS signal, a BER of 0.168 was observed. After applying monostatic cancellation, the BER decreased to 0.047. The estimated EVM also improved significantly, from \qty{0.85}{dB} to \qty{-6.23}{dB}.

\begin{figure}[!t]
	\begin{center}
		\includegraphics[width = 0.48\linewidth]{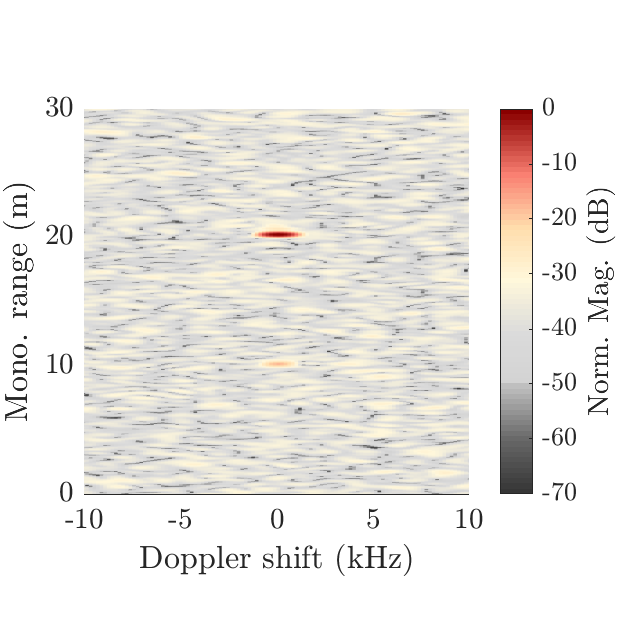}
		\includegraphics[width = 0.48\linewidth]{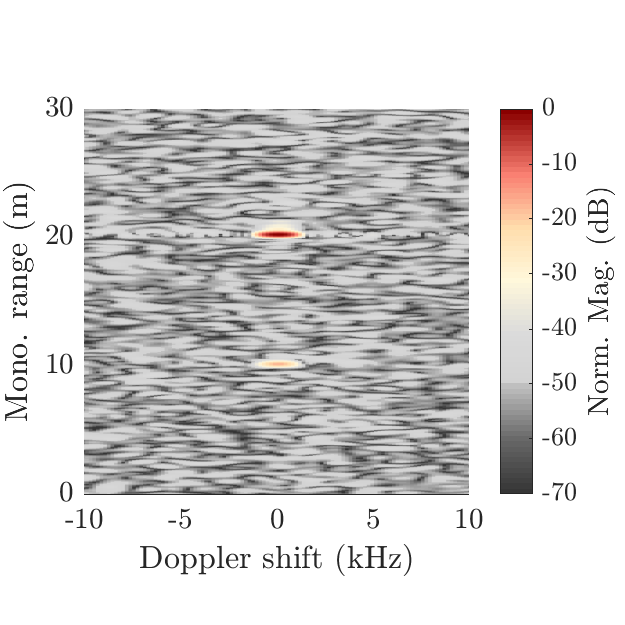}
		\\ (a) \qquad\qquad\qquad\qquad\qquad (b) 
		\caption{Measured monostatic RD maps: (a) before bistatic LoS cancellation, (b) after bistatic LoS cancellation.}\label{Fig18}
	\end{center}
\end{figure}

The results corresponding to the bistatic LoS cancellation case, where the estimated SINR was \qty{24}{dB}, are presented next. As shown in Fig.~\ref{Fig18}(a), the monostatic radar image is noticeably degraded due to interference from the bistatic LoS component, resulting in a measured image SINR of \qty{35.85}{dB}. Since this value is below the predefined threshold, bistatic LoS cancellation was applied. After the bistatic LoS cancellation, the monostatic radar image exhibits improvement as shown in Fig.~\ref{Fig18}(b). Specifically, with the image SINR increasing from \qty{37.79}{dB} to \qty{52.18}{dB}. In this case, the estimated BER and EVM were $4.9 \times 10^{-5}$ and \qty{-18.66}{dB}, respectively, indicating nearly error-free communications.

\begin{figure*}[!t]
	\begin{center}
		\includegraphics[width = 0.32\linewidth]{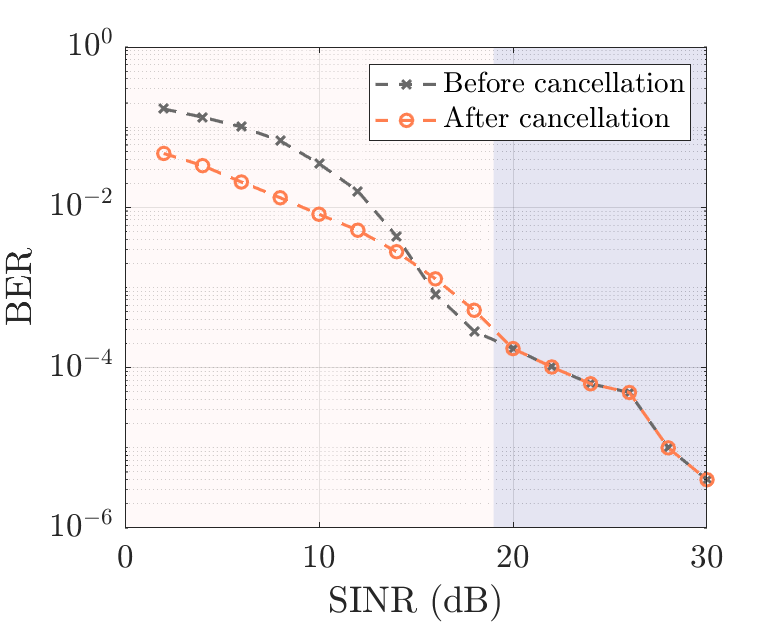}
		\includegraphics[width = 0.32\linewidth]{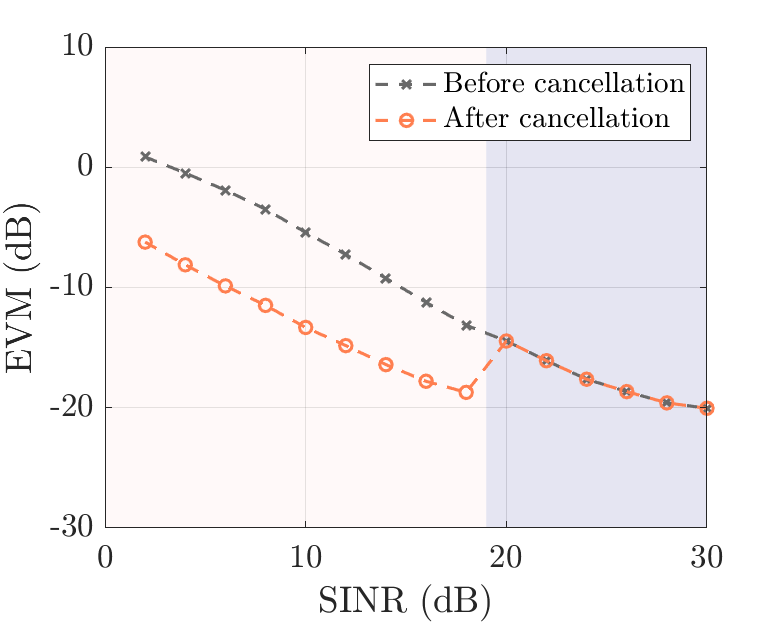}
		\includegraphics[width = 0.32\linewidth]{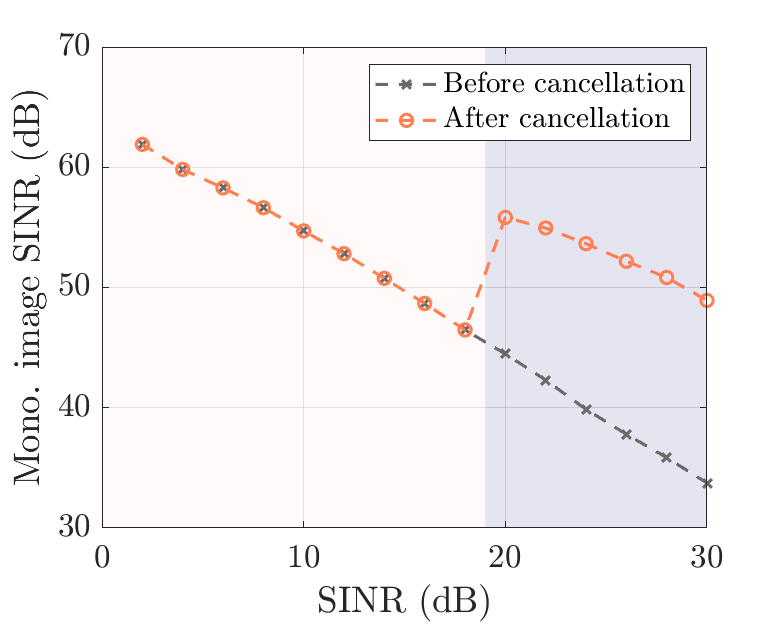}
		\\ (a) \qquad\qquad\qquad\qquad\qquad\qquad\qquad (b) \qquad\qquad\qquad\qquad\qquad\qquad\qquad (c) 
		\caption{Measurement-based comparison of sensing and communication performance before and after interference cancellation under varied SINR conditions: (a) BER, (b) EVM, and (c) monostatic radar image SINR.}\label{Fig19}
	\end{center}
\end{figure*}

Lastly, we varied the SINR value from \qty{0}{dB} to \qty{30}{dB} by adjusting the attenuation of the monostatic path. Fig.~\ref{Fig19} shows the comparison between before and after cancellation in terms of BER, EVM, and monostatic image SINR. As shown in Fig.~\ref{Fig19}(a), the BER continues to decrease as SINR increases for both before and after cancellation case. From \qty{0}{dB} to \qty{14}{dB} SINR, we can observe the decreased BER when cancellation is applied. Beyond \qty{14}{dB} SINR, the BER for both cases becomes nearly identical, which supports the fact that additional benefit w.r.t communications over \qty{14}{dB} SINR is highly limited as mentioned in Sec. \ref{Sec3_1}. However, unlike the simulation results, the BER does not converge to zero in measurements, due to the aforementioned hardware impairments. While forward error correction schemes like low-density parity-check codes can eliminate residual bit errors, such coding techniques were not considered in this work. Followed by the BER comparison, Fig.~\ref{Fig19}(b) shows the comparison of EVM. The average EVM gain achieved by the proposed cancellation was \qty{8.82}{dB}. The overall EVM curve for both before and after cancellation case shows a similar trend compared to the simulation results, except that both curves exhibit higher initial EVM values. This is because only AWGN was considered in simulations, resulting in less distortion compared to real-world scenarios. Finally, Fig.~\ref{Fig19}(c) shows the monostatic image SINR. Similar to the behavior of the EVM performance, measurement results with monostatic image SINR also tends to follow the simulation result, with the average gain in image SINR of \qty{13.59}{dB}.

\section{Conclusion}\label{Sec5}
This work investigated interference dynamics and proposed an interference cancellation strategy for multistatic OFDM-based ISAC systems. A scenario was considered in which a monostatic and a bistatic ISAC node co-exist and operate over the same spectral resources. The study demonstrated how the mono- and bistatic-oriented signals influence each other depending on their relative power levels at the receiver and evaluated the resulting performance degradation in both radar and communication processing. A cancellation method was proposed based on the monostatic radar image SINR, which serves as a decision metric to determine whether the monostatic reflection or the bistatic LoS component should be canceled. A heuristic threshold of \qty{45}{dB} was set to trigger the cancellation procedure. Simulation results showed that the proposed method could reduce the EVM by \qty{12.39}{dB} and enhance the monostatic radar image SINR by \qty{13.5}{dB}. Proof-of-concept measurements also confirmed the effectiveness of the proposed method, demonstrating an EVM reduction of \qty{8.82}{dB} and a SINR improvement of \qty{13.59}{dB}, closely matching the simulation results. While this work focuses on a simple configuration with a single monostatic and bistatic node, future research should extend the framework to scenarios with multiple distributed transmitters and receivers. In particular, adapting the approach to MIMO configurations and multi-transmitter networks will be essential to cope with the increased dimensionality and interference complexity in practical large-scale ISAC deployments.

\section*{Acknowledgments}
We would like to thank Rohde \& Schwarz for providing the radar echo generator ($\text{R\&S®}$ AREG800A).

\end{document}